\documentclass[12pt]{amsart}
\newcommand {\supplus}{\mathop{{\supset}\llap{\raise
0.5pt\hbox{\normalfont\small+}\hskip
0.5pt}}}

\newcommand {\subplus}{\mathop{{\subset}\llap{\raise
0.5pt\hbox{\normalfont\small+}\hskip 0.5pt}}}

\newcommand {\Cee}    {{\mathbb  C}}

\newcommand {\Ree}    {{\mathbb  R}}

\newcommand {\Zee}    {{\mathbb  Z}}

\newcommand {\fa}     {{\mathfrak{a}}}

\newcommand {\fab}    {{\mathfrak{ab}}}

\newcommand {\faut}   {{\mathfrak{aut}}}
\newcommand {\fb}     {{\mathfrak{b}}}
\newcommand {\fc}
{{\mathfrak{c}}}

\newcommand
{\fder}   {{\mathfrak{der}}}   %
\newcommand {\fdiff}  {{\mathfrak{diff}}}

\newcommand {\fg}     {{\mathfrak{g}}}
\newcommand {\fgl}    {{\mathfrak{gl}}}  %
\newcommand {\fh}     {{\mathfrak{h}}}
\newcommand {\fhei}   {{\mathfrak{hei}}}
\newcommand {\fk}     {{\mathfrak{k}}}

\newcommand {\fle}    {{\mathfrak{le}}}
\newcommand {\fm}     {{\mathfrak{m}}}

\newcommand {\fo}     {{\mathfrak{o}}}
\newcommand {\fosp}   {{\mathfrak{osp}}}
\newcommand {\fpe}    {{\mathfrak{pe}}}   %

\newcommand {\fpo}    {{\mathfrak{po}}}

\newcommand {\fq}     {{\mathfrak{q}}}

\newcommand {\fs}     {{\mathfrak{s}}}

\newcommand {\fsh}
{{\mathfrak{sh}}}

\newcommand {\fsl}    {{\mathfrak{sl}}}
\newcommand {\fsle}   {{\mathfrak{sle}}}
\newcommand {\fsm}
{{\mathfrak{sm}}}
\newcommand {\fsp}    {{\mathfrak{sp}}}
\newcommand {\fspe}   {{\mathfrak{spe}}}

\newcommand {\fsq}    {{\mathfrak{sq}}}

\newcommand {\fsvect} {{\mathfrak{svect}}}

\newcommand {\fvect}  {{\mathfrak{vect}}}   %

\newcommand {\cal}
{\mathcal}

\newcommand {\cC}     {{\cal C}}

\newcommand
{\cF}     {{\cal F}}

\newcommand {\cL}     {{\cal L}}
\newcommand {\cM}     {{\cal M}}

\newcommand {\cO}     {{\cal O}}

\def \opname#1#2%
  {\expandafter\newcommand \csname #1\endcsname {{\mathop{#2}\nolimits}}}

\newcommand{\rmname}[1]
  {\expandafter\newcommand \csname #1\endcsname {{\operatorname{#1}}}}

\newcommand{\rmnameii}[2]
{\expandafter\newcommand \csname #1\endcsname {{\operatorname{#2}}}}

\rmname{act}
\rmname{Ad}
\rmname{Add}
\rmname{ad}
\rmname{Alt}
\rmname{alt}
\rmname{Ann}
\rmname{antidiag}
\rmname{Ber}
\rmname{ber}
\rmname{Br}
\rmname{card}
\rmname{ch}
\rmname{Char}
\rmname{cem}
\rmname{cj}
\rmname{Cliff}
\rmname{cntr}
\rmname{codim}
\rmname{coind}
\rmname{const}
\rmname{col}
\rmname{cork}
\rmname{cpr}
\rmname{diag}
\rmnameii{Div}{div}
\rmname{Def}
\rmname{Der}
\rmname{Dim}
\rmname{End}
\rmname{Even}
\rmname{Ext}
\rmname{gr}
\rmname{Hom}
\rmname{HT}
\rmnameii{Ht}{ht}
\rmname{hwt}
\rmname{Id}
\rmname{id}
\rmname{ind}
\rmname{Ind}
\rmname{Inf}
\rmname{irr}
\rmname{Le}
\rmname{Lie}
\rmname{lwt}
\rmname{mult}
\rmname{Mor}
\rmname{nm}
\rmname{Ob}
\rmname{Odd}
\rmname{Osc}
\rmname{per}
\rmname{Pic}
\rmname{pr}
\rmname{pro}
\rmname{Prime}
\rmname{Proj}
\rmname{prt}
\rmname{pt}
\rmname{Q}
\rmname{qet}
\rmname{qtr}
\rmname{rd}
\rmname{rk}
\rmname{row}
\rmname{Res}
\rmname{salt}
\rmname{Sch}
\rmname{SBr}
\rmname{scalar}
\rmname{Ser}
\rmname{sign}
\rmname{Smbl}
\rmname{spin}
\rmname{ssym}
\rmname{str}
\rmname{st}
\rmname{sgn}
\rmname{sq}
\rmname{symm}
\rmname{supp}
\rmname{Supp}
\rmname{St}
\rmname{Spec}
\rmname{Spm}
\rmname{tr}
\rmname{vpt}
\rmname{weyl}
\rmname{Weyl}
\rmname{Witt}

\opname{vvol}  {{v\hspace{-0.1ex}o\hspace{-0.02ex}l\/}}
\opname{pnt}  {\text{\normalfont
pt}}
\opname{Span} {{Span}}
\opname{slim} {\overline{\lim}}
\opname{Vol}  {{V\hspace{-0.55ex}o\hspace{-0.02ex}l\/}}
\opname{Par}  {{P\hspace{-0.3ex}a\hspace{-0.05ex}r\/}}

\rmname{Mat}
\rmname{Bil}
\rmname{Diff}
\rmname{Ker}
\rmname{Herm}
\rmname{Coker}
\rmname{Conn}
\rmname{Covect}
\rmname{Vect}
\rmname{Int}

\rmnameii {IM} {Im}
\rmnameii {RE} {Re}

\opname{Aut}
{{A\hspace{-0.2ex}u\hspace{-0.1ex}t\/}}
\opname{GL} {{G\hspace{-0.3ex}L}}
\opname{SL} {{S\hspace{-0.3ex}L}}
\opname{Exp} {{E\hspace{-0.2ex}x\hspace{-0.1ex}p\/}}
\opname{GQ} {{G\hspace{-0.2ex}Q}}
\opname{OSp}
{{O\hspace{-0.25ex}S\hspace{-0.15ex}p\/}}
\opname{Out} {{O\hspace{-0.25ex}u\hspace{-0.15ex}t\/}}
\opname{Spp} {{S\hspace{-0.2ex}p\/}}
\opname{SpO}
{{S\hspace{-0.2ex}p\hspace{-0.02ex}O\/}}
\opname{Pe} {{P\hspace{-0.25ex}e\/}}
\opname{SPe} {{S\hspace{-0.25ex}P\hspace{-0.25ex}e\/}}
\opname{Spin} {{S\hspace{-0.25ex}p\hspace{-0.05ex}i\hspace{-0.1ex}n\/}}
\opname{Iso}
{{I\hspace{-0.25ex}s\hspace{-0.1ex}o\/}}
\opname{SSPe} {{S\hspace{-0.25ex}S\hspace{-0.15ex}P\hspace{-0.25ex}e\/}}
\opname{PeU} {{P\hspace{-0.25ex}e\hspace{-0.1ex}U\/}}
\opname{QU}
{{Q\hspace{-0.15ex}U\/}}
\opname{U} {{U\/}}

\opname{cGQ} {{\cal G \hspace{-0.2em} Q \/}}
\opname{cSL} {{\cal S \hspace{-0.2em} L \/}}
\opname{cGL} {{\cal G \hspace{-0.2em} L
\/}}
\opname{cOSp} {{\cal O \hspace{-0.2em} S \hspace{-0.3em} \it p\/}}
\opname{cPe} {{\cal P \hspace{-1.5pt} \it e\/}}
\opname{cVect} {{\cal V \hspace{-1.5pt} \it
e\hspace{-0.1ex}c\hspace{-0.1ex}t\/}}
\opname{cVol} {{\cal V \hspace{-1.5pt} \it o\hspace{-0.1ex}l\/}}
\opname{cAut} {{\cal A \hspace{-0.2em} \it u\hspace{-0.1em}t\/}}
\opname{cCovect} {{\cal C
\hspace{-1.5pt}
     \it o\hspace{-0.1ex}v\hspace{-0.1ex}e\hspace{-0.1ex}c\hspace{-0.1ex}t\/}}
\opname{CW} {{C\hspace{-0.15ex}W}}

\newcommand {\ev}
{{\bar0}}
\newcommand {\od} {{\bar1}}
\newcommand {\eps} {\varepsilon}
\newcommand {\degree}  {{}^\circ}
\newcommand {\tto} {\longrightarrow}
\newcommand {\pder}[1]
{{\frac{\partial}{\partial {#1}}}}
\newcommand {\pderf}[2] {{\frac{\partial {#1}}{\partial {#2}}}}

\newcommand {\bcdot}   {\mathbin{\hbox{\raise.4ex\hbox{\bf.}}}} 

\newcommand {\secno} {}
\newcommand
{\ssecfont} {\normalfont\bf}

\newtheorem*{Theorem}{\secno Theorem}

\newenvironment {th*}[1]
    {\gdef\thname{#1} \begin{thn}}%
    {\end{thn}}
\newtheorem*{thn} {\thname}

\theoremstyle{definition}

\newenvironment {ex*}[1]
    {\gdef\thname{#1}
\begin{exn}}%
    {\end{exn}}
\newtheorem*{exn}{\thname}

\theoremstyle{remark}

\newtheorem*{Remark}{\secno Remark}
\newtheorem*{Remarks}{\secno Remarks}

\newenvironment {rem*}[1]
    {\gdef\thname{#1} \begin{remn}}%
{\end{remn}}
\newtheorem*{remn}{\thname}

\newcommand
{\ssec}{\subsection*}

\newcommand {\ssbegin}[2]
  {\def \secno {\gdef \secno {}{\ssecfont #1. }}
\begin{#2}}

\begin{document}

\title{How to quantize the antibracket}

\author{D. Leites${}^1$ and I. Shchepochkina${}^2$}

\address{${}^1$Department of Mathematics, University of Stockholm,
Roslagsv. 101, Kr\"aftriket hus 6, SE-104 05, Stockholm, Sweden,
e-mail: mleites@math.su.se; ${}^2$Independent University of
Moscow, Bolshoj Vlasievsky per., RU-121 002 Moscow, Russia;
irina@mccme.ru}



\begin{abstract} The
uniqueness of (the class of) deformation of
Poisson Lie algebra $\fpo(2n)$ has long been a completely accepted
folklore.  Actually this is wrong as stated, because its
validity
depends on the class of functions that generate $\fpo(2n)$ (e.g.,
it is true for polynomials but false for Laurent polynomials).

We show that, unlike $\fpo(2n|m)$, its quotient
modulo center, the
Lie superalgebra $\fh(2n|m)$ of Hamiltonian vector fields with
polynomial coefficients, has exceptional extra deformations for
$(2n|m)=(2|2)$ and only in this
superdimension.  We relate this
result to the complete description of deformations of the
antibracket (also called the Schouten or Buttin bracket).

We show that, whereas the
representation of the deform (the result
of deformation aka quantization) of the Poisson algebra in the
Fock space coincides with the simplest space on which the Lie
algebra of commutation
relations acts, this coincidence is not
necessary for Lie superalgebras.

\end{abstract}

\maketitle

\section{Introduction}

This is an edited version of the paper preprinted in Erwin
Schr\"odinger International Institute for Mathematical Physics
(875; www.esi.ac.at) and published in Theor.  Math. Phys.  To save
space, and having in mind most general target audience, mainly
interested in answers, we have omitted boring calculations
(including an exposition of important but inaccessible paper
\cite{Ko2}). The omitted material whose documentation took far too
long time will be published together with the details of the proof
of our classification of simple vectorial Lie superalgebras: an
expounding of \cite{LS2}. As usual, when one deletes something
``obvious'' one should be extra careful and we are sorry to say
that we did throw away several cocycles (fortunately, on
isomorphic algebras). We also modify the final text by disclaiming
Shmelev's interpretation of $\fh_{\lambda}(2|2)$ which we used to
trustfully rewrite from paper to paper.

We compensate the proofs omitted by extensive background:
several
vital, not just important, notions (for example, that of Lie
superalgebra) are not as well known as is the general belief.

\ssec{1.1.  General setting of our
problem} The
problem we consider is usually breezily formulated.
In order not to get confused and derive our main result, we do our
best to formulate it extra carefully.

In 1977, M.~Marinov asked
one of us: ``How to quantize this ``new
mechanic (\cite{L1})'' of yours?  Will the Planck's constant be
odd?!''  In 1987, S.~Sternberg repeated the question in connection
with his
studies with Kostant \cite{KS}.  For a preliminary
answer, see \cite{L3}, where the importance of odd parameters and
the ``queer'' analog of $\fgl$ was indicated.  Here we
concentrate
on other issues but again, as in \cite{L3}, tirelessly emphasize
the importance of the ``point functor'' approach to Lie
superalgebras. In particular, if we deal with their
deformations
one needs odd parameters.  For the convenience of the reader, all
necessary background is collected in \S 4 (Background).

The main questions, before we start counting how
many
quantizations of the Poisson algebra are possible, are {\bf what
is quantization and what is the Poisson algebra?}

There are many interpretations of the notion
``quantization''. We
consider quantization as a deformation, and it is vital to start
with a lucid description of the class in which we deform our
object, to say nothing of the lucid
description of the object
itself. For example, in the simplest case, when the supermanifold
$\cM$ is $\Cee^{2n|m}$ (or $\Ree^{2n|m}$) equipped with a
symplectic structure, we consider the
superspace $\cF$ of
functions on $\cM$. There are two natural structures on $\cF$:
that of an {\it associative} (and supercommutative) superalgebra
and that of a {\it Lie}
superalgebra, called the {\it Poisson
superalgebra} and denoted by $\fpo(2n|m)$. Therefore we must first
select one of the two problems: describe either

\medskip

(1) deformation of the
{\it associative} superalgebra $\cF$ (usually, one
sacrifices commutativity) or

(2) deformation of the {\it Lie} superalgebra $\fpo(2n|m)$.

\medskip

\noindent Both problems can be
solved by computing a certain
cohomology (Hochschield one for deformations of the associative
algebra structure, Lie one for deformations of the Lie algebra
structure; passage to
superagebras only brings in some extra
signs). Problem (1) was considered from various angles by Flato
et. al. \cite{Bea}, Neroslavsky and Vlassov \cite{NV}, De Wilde
and Lecomte,
Drinfeld, Fedosov (\cite{Fe}, \cite{D1}), Kontsevich
\cite{Kon1} to name a few.

It was always clear that Problems (1) and (2) are related; here we
intend not to replace one with the
other but manifestly separate
them and concentrate on Problem (2).  Dirac was, perhaps, the
first to consider it: indeed, quantization in \cite{Dir} is
understood as follows. Let
$(M, \omega)$ be a symplectic manifold
(locally: domain), $\{\cdot , \cdot\}$ the corresponding Poisson
bracket. Assume that all functions (\lq\lq observables") depend on
a
parameter $t$ (time). Then {\bf quantization} is a passage from
the classical equations of motion with Hamiltonian $H$
\begin{equation}\label{cem}
\dot{f}=\{f, H\}
\end{equation}
to
quantum ones
\begin{equation}\label{qem}
\dot{\hat f}=[\hat f, \hat H],
\end{equation}
where $\hat f$ and $\hat H$ are operators (acting in a space to be
specified) and $[\cdot , \cdot]$ is the commutator. Let us express
the elements of the Poisson algebra (the product in which
is$\{\cdot , \cdot\}$) as contact fields $K_f$ with generating
function $f$ (see (\ref{eq56})), so that eq. (\ref{cem}) becomes
\begin{equation}\label{cemkf}
\dot{K_f}=[K_f, K_H].
\end{equation}
In this formulation, it becomes manifest that the structure of
associative and commutative
algebra on the space of functions
which label the classical operators $K_f$ is beside the point;
whereas quantization is, equally manifestly, a deformation of the
Lie algebra
structure inside the variety of Lie algebras.

For a long time Vey's paper \cite{V} was the only one where the
deformation of {\it Lie} structure was studied (cf. \cite{HG});
in
\cite{L3} and here we follow this approach.

Observe a totally different from anyone's approach to quantization
due to Berezin \cite{B}, where the dimensions of the algebras
deformed (in
Berezin's sense) can vary under deformation and where
the convergence of the series expansion of the formal parametric
family of multiplications is investigated.

Concerning Problem (1), Shereshevskii \cite{SI} was the first, as
far as we know, to show that the space of deformations of the
associative structure on $\cF(\cM)$ is \lq\lq not less", in a
sense, than the space of affine connections on $\cM$ and is,
therefore, too huge to be of interest (is undescribable).
Kontsevich \cite{Kon1} understood that this space should be
considered modulo certain gauge transformations and showed that
this makes the quotient space describable (one dimensional).
Superization of this result is a routine job performed in
\cite{Bo}.

In
pre-Kontsevich era, to diminish the number of deformations of
the {\it associative} structure, people usually assumed that the
following \lq\lq correspondence principle" holds
\begin{equation}
\label{eq1}
\lim_{\hbar\tto 0}\frac{f*_{\hbar}g -g*_{\hbar}f}{\hbar}=\{f, g\}.
\end{equation}
This, actually, amounts to replacement of Problem
(2) by
Problem (1).

\begin{Remark} Observe that Kontsevich even considered the Lie bracket
constructed from not necessarily
non-degenerate odd bivector
field. The bracket thus obtained is sometimes also called Poisson
bracket augmenting the already considerable confusion.
\end{Remark}

\ssec{1.2.
Cohomology depend on the type of functions} The number
of nonequivalent deformations of the Lie (super)algebra $\fg$ may also
depend on the type of functions involved in the
description of $\fg$
(smooth, analytic, polynomial, etc.): compare the parametric family
$\fsvect_{\lambda}^L(1|n)$ of divergence free vector fields with
Laurent polynomials as coefficients
\cite{GLS}, with the rigid Lie
superalgebra $\fsvect(1|n)$ of divergence free vector fields with
polynomial coefficients.

The uniqueness of deformation of $\fpo(2n)$ was a
folklore since
long ago.  When Batalin and Tyutin \cite{BT1} actually {\it
proved} the statement for Poisson Lie superalgebras $\fpo(2n|m)$
(generated by functions of a certain class),
they found it
difficult to publish the result because it was dubbed as ``known''
(although no {\it proof} was ever published even in the purely
even case, cf.  \cite{V}, review
\cite{Bea} and more recent
\cite{HG}, the object under study being the existence, not
uniqueness).  However, one should be very careful here: for
arbitrary generating functions, the
statement is wrong, because
its validity depends on the class of functions: For example, for
polynomials this is true, but false for Laurent polynomials, cf.
\cite{Dzh}, see also \cite{KT}. Another
example is a multiparameter quantization of
functions on the orbits of simple Lie groups in the coadjoint
representation, cf. \cite{DGS}, \cite{Kon2}, \cite{GL2}.

\ssec{1.3.  What a Lie
superalgebra is} Lie superalgebras had
appeared in topology in 1930's or earlier.  So when somebody
offers a ``better than usual'' definition of a notion which seemed
to have been
established about 70 year ago this might look
strange, to say the least.  Nevertheless, the answer to the
question ``what is a Lie superalgebra?''  is still not a common
knowledge.
Indeed, the naive definition (``apply the Sign Rule to
the definition of the Lie algebra'') is manifestly inadequate for
considering the (singular) supervarieties of deformations
and
applying representation theory to mathematical physics, for
example, in the study of the coadjoint representation of the Lie
supergroup which can act on a supermanifold but never on
a
superspace (an object from another category).  So, to deform Lie
superalgebras, apply group-theoretical methods in ``super''
setting, etc., we must be able to recover a
supermanifold from a
superspace, and vice versa.

A proper definition of Lie superalgebras is as follows, cf.
\cite{L3}.  The {\it Lie superalgebra} in the category of
supermanifolds
corresponding to the ``naive'' Lie superalgebra $L=
L_{\ev} \oplus L_{\od}$ is a linear supermanifold $\cL=(L_{\ev},
\cO)$, where the sheaf of functions $\cO$ consists of functions
on
$L_{\ev}$ with values in the Grassmann superalgebra on $L_{\od}^*$;
this supermanifold should be such that for \lq\lq any" (say, finitely
generated, or from some other appropriate
category) supercommutative
superalgebra $C$, the space $\cL(C)=\Hom (\Spec C, \cL)$, called {\it
the space of $C$-points of} $\cL$, is a Lie algebra and the
correspondence $C\tto
\cL(C)$ is a functor in $C$.  (A.~Weil
introduced this approach in algebraic geometry in 1953; in super
setting it is called {\it the language of points} or {\it families},
see
\cite{L}.)  This definition might look terribly complicated, but
fortunately one can show that the correspondence
$\cL\longleftrightarrow L$ is one-to-one and the Lie algebra
$\cL(C)$,
also denoted $L(C)$, admits a very simple description: $L(C)=(L\otimes
C)_{\ev}$.

A {\it Lie superalgebra homomorphism} $\rho: L_1 \tto L_2$ in these
terms is a functor morphism,
i.e., a collection of Lie algebra
homomorphisms $\rho_C: L_1 (C)\tto L_2(C)$ compatible with morphisms
of supercommutative superalgebras $C\tto C'$.  In particular, a
{\it
representation} of a Lie superalgebra $L$ in a superspace $V$ is a
homomorphism $\rho: L\tto \fgl (V)$, i.e., a collection of Lie algebra
homomorphisms $\rho_C: L(C) \tto ( \fgl (V )\otimes
C)_{\ev}$.

\begin{rem*}{Example} Consider a representation $\rho:\fg\tto\fgl(V)$.
The tangent space of the moduli superspace of deformations of $\rho$
is isomorphic to $H^1(\fg;
V\otimes V^*)$.  For example, if $\fg$ is
the $0|n$-dimensional (i.e., purely odd) Lie superalgebra (with the
only bracket possible: identically equal to zero), its
only
irreducible representations are the trivial one, {\bf 1}, and
$\Pi({\bf 1})$.  Clearly, ${\bf 1}\otimes {\bf 1}^*\simeq \Pi({\bf
1})\otimes \Pi({\bf 1})^*\simeq {\bf 1}$, and because the
superalgebra
is commutative, the differential in the cochain complex is trivial.
Therefore $H^1(\fg; {\bf 1})=E^1(\fg^*)\simeq\fg^*$, so there are
$\dim\,\fg$ odd parameters of
deformations of the trivial
representation.  If we consider $\fg$ ``naively'' all of the odd
parameters will be lost.

Which of these infinitesimal deformations can be
extended to a
global one is a separate much tougher question, usually
solved {\it ad hoc}, see \cite{F}. \end{rem*}

In this paper we deal with a similar problem: we deform the Lie
superalgebra
structure, i.e., the superbracket.  Deformations of any
superstructure can, of course, have odd parameters.  Yu.~Manin writes
that this is obvious \cite{Man} but he is overoptimistic:
even live
classics still sometimes deliberately ignore odd parameters, see,
e.g., \cite{CK}.  Physicists easier accept odd (and other
infinitesimal) parameters; odd parameters are
the cornerstone of
supersymmetry (\cite{WZ}); in several famous papers Witten clearly
illustrated the importance of odd parameters, see, for example,
\cite{W}.  Witten's papers
triggered an avalanche of elaborations
among which we would like to point out \cite{GK}, \cite{Man},
\cite{R}, \cite{Shu}.

\ssec{1.4.  Quantization, as we understand it} Quantization
of
$\fpo(2n|m)$ consists of two steps:

\medskip

(1) deformation of the Lie superalgebra $\fpo(2n|m)$ and

(2) realization of the deform by operators in some space (the
Fock
space).

\noindent There is also a step somewhat aside, 0-th step:

(0) prequantization, i.e., realization of $\fpo(2n|m)$ by operators in
some ``classical version'' of the Fock space.

\medskip

Execution of Steps (1) and (2) seems to be routine; their
superization has only two novel features: realization of
$\fpo(2n|m)$ for $m$ odd not by all differential operators but by
a part similar to the ``queer'' analog of the general Lie algebra,
the one which preserves a complex structure given by an odd
operator.  For details, see \cite{L3}.  Another novel feature is
described in sec.~1.5.

For a complete description of prequantizations, see \cite{BSS},
\cite{Sm1}, \cite{Ko6}.

Related with the prequantization is description of
representations
of (anti)commutation relations (RCR).  (For an approach distinct
from ours, see \cite{B1}.)  It turns out that the representation
of the deform (after quantization) of
$\fpo(2n|0)$ in the Fock
space coincides with the simplest space on which RCR act.  Observe
that this coincidence is not necessary for Lie superalgebras.

Here we consider the odd
analogs of the Poisson bracket, namely,
the antibracket (Schouten or Buttin bracket) and the deformations
of the antiblracket, other than quantizations.  For each of these
Lie
superalgebras with these brackets, i.e., for $\fpo(2m|n)$,
$\fb(n)$ and each member $\fb_\lambda(n)$ of the the one-parameter
set of deformations of $\fb(n)$, we will investigate
its
quantization in the above sense.

Speaking about antibracket, recall that the {\it Buttin superalgebra}
$\fb(n)$ is the superspace of functions with reversed parity on
the
$n|n$-dimensional superspace endowed with the Lie superalgebra
structure given by the antibracket; $\fb(n)$ can also be realized as
the superspace of multivector fields (with reversed
parity) on
$p|q$-dimensional superspace for any $p$, $q$ such that $p+q=n$, $p,
q\geq 0$ and with the Lie superalgebra structure given by the Schouten
bracket.

\begin{Remark} 1) What we
call the ``Buttin bracket'' here was discovered in the pre-super
era by Schouten; Buttin first proved that this bracket establishes
a Lie superalgebra structure. The interpretations of the Buttin
superalgebra $\fb(n)$ similar to that of the Poisson algebra
$\fpo(2n|m)$ and of the elements of $\fle(n)=\fb(n)/\text{center}$
as analogs of Hamiltonian vector fields was given in \cite{L1}.
The Buttin bracket and ``odd mechanics'' introduced in \cite{L1}
was rediscovered by Batalin and Vilkovisky (and, even earlier, by
Zinn-Justin, but his papers went mainly unnoticed, as observed in
\cite{FLSf}); it gained a great deal of currency under the name
{\it antibracket}; in several papers Batalin and Vilkovisky
demonstrated its importance, see reviews \cite{GPS}, \cite{BT2}.
\end{Remark}

{\sl Not every deformation qualifies to be considered as
quantization.} Roughly speaking, having started with a Lie
(super)algebra of vector fields on a superspace of certain
dimension, we should, after quantization, obtain an algebra which
possesses a representation in the space of halved functional
dimension.  The deforms (results of the deformation) of $\fb(n)$
given by (\ref{eq7}) are denoted by $\fb_{\lambda}(n)$, see
Background.  NONE of the deformations of $\fb(n)$ is quantizations
in the above sense (there are no representations of halved
dimension).  The one we call quantization just looks similar to
the only quantization of Poisson algebra.

For the odd versions of prequantizations, i.e., representations of
$\fb_{\lambda}(n)$,
and related with them description of the
representations of $\fle(n)$, see \cite{L2}, \cite{Ko6}.

Step (1) was performed by Kochetkov in \cite{Ko1}--\cite{Ko5}
(except for a case missed; we will also show that this omission is
inessential) for the Buttin superalgebras $\fb_{\lambda}(n)$, and
for $\fh(2n|m)$ except for $nm\neq 0$. Here we correct Kochetkov's
result and complete Step (2) started in \cite{L3}.

\ssec{1.5.  Numerous Fock spaces} There are two major types of the new
Fock spaces:

1) Fix a realization (grading) of $\fpo$ or $\fb$.  Then $\fpo_-$ and
$\fb_-$ can
be considered as the analog of the Heisenberg Lie
superalgebra of anti/commutation relations.  The relatively new
message is that while for $\fpo_-$ there is only one (up to
a
character and parity change) irreducible representation, {\it there
are several non-isomorphic irreducible representations for the
(anti)commutation relations represented by
$\fb_-$.}

Throughout the paper we insist on considering $\fpo$ and its ``odd''
analog, $\fb$, as well as the deforms of the latter, $\fb_{\lambda}$,
as Lie superalgebras.  These Lie
superalgebras, and especially their
quantum analogs, are, in a sense, analogs of $\fgl(V)$.  The Lie
algebra $\fgl(V)$ has many irreducible representations (realized in
tensors
constructed on $V$ and $V^*$, in modules with vacuum vector,
etc.), and so do all the above mentioned Lie superalgebras.

Contrariwise, the associative algebra $\Mat(V)$ of endomorphisms
of
$V$ or its matrix version, $\Mat(\dim V)$, though isomorphic to
$\fgl(V)$ as a vector space, has only one irreducible module, $V$
(and so do the super versions of $\Mat(V)$, even
the queer
versions, $Q(V)$).  This module $V$ is exactly what is called the
{\it Fock space}.  So the Fock space is the analog of the standard
or identity representation for
$\fgl(V)$.  Therefore, considering
representations of the {\it Lie} (super)algebra, we should take
the ``smallest'' representation, the one which plays the role of
the identity one for
$\fgl(V)$ for the role of an analog of the
Fock space, especially in the ``classical'' case, that of $\fpo$
or $\fb$.

2)  {\it Every ``nonstandard realization'' of $\fpo$ and $\fb$ has
its own Fock space; several ones in case of $\fb$}. Here we draw
attention of the reader to the following phenomenon.  Even for the
``conventional'' Poisson {\it super}algebra $\fpo(2n|m)$, there
are {\it several} analogs of the Fock space representations
corresponding to several {\it nonstandard realizations}
$\fpo(2n|m; r)$ of $\fpo(2n|m)$.  For example, these realizations
for $m=2k$ are given by the following gradings of the generating
functions $p, q, \xi, \eta$:
\begin{equation}
\label{eq2}
\renewcommand{\arraystretch}{1.4}
\begin{array}{l}
    \deg
p_{i}=\deg q_{i}=\deg
\xi_{j}=\deg \eta_{j}=1\text{ for any $i$ and }j>r; \\
\deg \xi_{j}=0,\; \deg \eta_{j}=2 \text{ for } 0\leq j\leq r.
\end{array}
\end{equation}
For the complete list of nonstandard realizations --- one of the
main results of classification of simple vectorial Lie
superalgebras, see \cite{Sch}, \cite{LS1}.  We only need some of
them, see sec.~A.7. Here we describe in detail only one of these
realizations, the standard one (Theorem 3.2).

Observe that though for distinct nonstandard realization the
analogs of $\fhei$ and $\fab$ are
of different dimensions, the
adjoint representations of $\fpo$ and $\fb$ are the ``smallest''
ones and, in contradistinction with numerous analogs of Fock
spaces for representations
of (anti)com\-mu\-ta\-tion relations,
are unique.

\ssec{1.6.  Main results} 1)  We observe that the Lie
superalgebras $\fh(2|2)$ of Hamiltonian vector fields can be
included into a parametric family $\fh_{\lambda}(2|2)$.  Theorem
2.1 shows how deformations of the Lie superalgebra structures
given by the Poisson bracket and antibracket are interrelated,
namely, we show that $\fh_{\lambda}(2|2)$ is isomorphic to the
regrading $\fb_{\lambda}(2; 2)$ of $\fb_{\lambda}(2)$.

From here we deduce new exceptional quantizations of the
Lie
superalgebras $\fh(2|2)$ of Hamiltonian vector fields and the Buttin
superalgebra $\fb(2)$.

2)  As a corollary of the above we deduce that {\bf the Lie
superalgebra of Hamiltonian vector fields may have more
quantizations than the corresponding Poisson one}.  It seemed
natural to expect that the Lie superalgebra $\fh(2n|m)$ of
Hamiltonian vector fields --- the quotient of $\fpo(2n|m)$ modulo
center --- has exactly one quantization, as many as $\fpo(2n|m)$.
These great expectations are justified almost always, except for
$(2n|m)=(2|2)$, when the parameters of deformation belong to a
singular supervariety {\it almost} completely described by
Kochetkov \cite{Ko1}--\cite{Ko5}.  His omission should have been
obvious in view of our earlier result \cite{ALSh} but everybody
overlooked it; perhaps, because it does not actually matter.  Here
we conclude the description of the deformations of $\fh(2n|m)$ and
relate them with the complete description of quantizations
(deformations) of the antibracket and its quotient modulo center,
$\fle(n)$.

3)  Unlike the quantized Poisson algebra $\fpo(2n|0)$ and
$\fhei(2n|0)$ which have exactly ONE realization by means of
creation and annihilation operators in the {\it Fock space}, the
quantized Lie superalgebra $\fab(n)$, the ``odd'' version of
$\fhei(2n|m)$, has $n+1$ distinct Fock spaces, one of which is of
finite dimension.  An important feature here: odd parameters of
representations are a must.

\begin{Remarks} 1)  Kochetkov proved
\cite{Ko2} that the subalgebra $\fs\fb(n)$ of divergence-free
multivector fields (superfields harmonic with respect to the odd
Laplacian) is not rigid and described the corresponding cocycles
in \cite{Ko2}, see also more accessible \cite{L3}.  The main
deformation of $\fb(n)$ described below preserves $\fs\fb(n)$;
this means that deformations of $\fs\fb(n)$ are of a different
nature. Still, the restriction of the quantization onto
$\fs\fb(n)$ is nontrivial; note that Kochetkov showed that there
are also {\it other} deformations of $\fs\fb(n)$.

2)  Our second main result shows that the Lie superalgebra of
Hamiltonian vector fields may have more quantizations than the
corresponding Poisson one just once: in dimension $(2|2)$.
Contrariwise, $\fle(n)$, the Lie superalgebra analogous to
$\fh(2n|m)$, is always more rigid than $\fb(n)$: namely, the
quantization does induce a deformation of $\fle(n)$, but that is
all: $\fle(n)$ has no other deformations.

3)  The passage to real forms is always possible whereas
exposition and study are easier over $\Cee$.  So in what follows
we work over $\Cee$.  (Passage from $\Cee$ to $\Ree$ should be
performed with caution: compare ``Theorem'' 9 of \cite{K} with
correct results of M.~Parker and Serganova \cite{S} and with
\cite{LS2}.)
\end{Remarks}

Note that using theorems from \cite{F} the volume of calculations
can be reduced to a negligible amount in the contact case as well
(sec.~3.3).  These simplifications are applicable to vectorial Lie
superalgebras with {\it polynomial} coefficients.  The case of
Laurent coefficients, especially for centrally extended algebras,
is quite different technically (or at least so it looks to us at
the moment); for partial results, see \cite{Ko4}.

\ssec{1.7.  On two confusions} 1)  The tendency to mix the
elements of the {\it Lie} superalgebra $\fpo(2n|m)$ labelled by
functions with the functions themselves (that generate an
associative and supercommutative superalgebra with respect to the
dot product) introduces a mess and hinders the study of {\it
quantization} in our sense, i.e., deformation of $\fpo(2n|m)$ as a
{\it Lie} superalgebra.

To
emphasize the distinction, we will denote the associative
(super)algebras by Latin characters (say, $A$); the same space
considered as a Lie (super)algebra with the (super)bracket $[x,
y]$
instead of the dot product $xy$ will be denoted by the corresponding
Gothic letter ($\fa$) or subscript $L$ for Lie ($A_{L}$).

2)  The situation is further worsened by the ``common knowledge''
of the following ``fact'' (see, e.g., Remark on p.  66 in Kac's
paper \cite{K}):
\begin{equation}
\label{eq3}
\text{ {\sl there exists an associative superalgebra $A$ such
that
$A_{L}\simeq\fpo(2n|m)$.}}
\end{equation}

\noindent This statement is wrong.  We will explain why and
eventually give a correct formulation, but first consider an
example which illuminates the problem.

In textbooks and papers (see, e.g., \cite{Pe}) the following
description of $\fpo(2n)$ can be encountered:

\medskip
{\sl $\fpo(2n)$ is generated }(presumably, as an
associative
algebra, whereas in fact it is a Lie algebra; for its presentation
as a {\it Lie algebra} in terms of generators and relations, see
\cite{LP}) {\sl by the $p_{i}$ and
$q_{i}$ for $i=1$, \dots $n$
subject to relations
\begin{equation}
\label{eq4}
\{p_{i}, q_{j}\}=i\hbar\delta_{ij}
\end{equation}
and the bracket should satisfy the Leibniz rule:}
\begin{equation}
\label{eq5}
\{f, gh\}=\{f, g\}h+g\{f,
h\}.
\end{equation}

\medskip

Obviously, Eq.  (\ref{eq5}) is not part of the definition of $\fpo(2n)$
but one of its properties, a particular case of the Lie derivative
along the vector
field generated by $f$, see \cite{BSS}.  (Eq.
(\ref{eq5}) is, however, a part of a definition of a {\it generalized}
Poisson structure, the one determined by a {\it degenerate}
bivector, as,
e.g., in \cite{Kon1}.)

Eqs.  (\ref{eq4}) are identities that determine the {\it Heisenberg Lie}
algebra $\fhei(2n)$ whose space is a $(2n+1)$-dimensional space
$W\oplus \Cee z$, where
$W$, spanned by $p, q$, is endowed with
the non-degenerate skew-symmetric form $B$, and $z$ lies in the
center and the Lie bracket is given by (\ref{eq4}) with the right hand
side
multiplied by $z$.

Our nihilistic stand towards associative algebras is justified (we
hope) by our results and some clarification of the general
picture. But in other problems one {\it
has} to consider both
structures together.  In his studies of integrable systems
Drinfeld even introduced the notion of {\it Poisson--Lie} algebra
(with both an associative and Lie
multiplications related by
Leibniz rule).  Here we do not consider this notion.

\ssec{1.7.1} From $\fhei(2n)$ we construct $\fpo(2n)$ in two steps.

{\bf Step 1}.  We consider the {\it
associative} algebra
$Weyl(2n)=U(\fhei(2n))$.  Because we are interested, mostly or only,
in {\it irreducible} representations of $\fhei(2n)$, we recall Schur's
lemma and fix the
central charge, rather than consider it a parameter,
i.e., identify $z$ with $i\hbar$.  The associative algebra
$\text{diff}(n)$ of differential operators with
polynomial
coefficients on an $n$-dimensional space can be viewed as
$U(\fhei(2n))/(z-i\hbar)$.  Both $Weyl(2n)$, and its quotient
$\text{diff}(n)$ are often called the {\it Weyl algebra}, from
the
context one can usually guess which of the two is meant.

{\bf Step 2}.  The Poisson algebra is not isomorphic to
$\fdiff(2n)=\text{diff}(2n)_{L}$ but is obtained from $\fdiff(2n)$
by
contraction, i.e., the passage to the quasi-classical limit as
$\hbar\tto 0$ after we set $p=i\hbar\pder{q}$ in (0.3).

Alternatively, one can define the Poisson algebra as isomorphic
to
$\gr(\fdiff(2n))$, the {\it graded} Lie algebra associated with
filtration of $\fdiff(2n)$ induced by the natural filtration of the
enveloping algebra $U(\fhei(2n))$.

\ssec{1.8.
Related problems} 1)  Having established the uniqueness of
the quantization, it is desirable to have a regular procedure for it.
On the flat space, there are several ways to pass
from the function
(i.e., the symbol, generating an element of the Poisson algebra) to
the corresponding operator; these procedures are Weyl, Wick, $pq$,
etc., quantizations.  The
uniqueness theorem states that all of them
are essentially equivalent.  Description of quantizations on the
spaces locally equivalent (in terms of $G$-structures) to
classical
domains had been started only recently, see \cite{DLO}, \cite{LO}.

2)  For {\it presentation} (i.e., description of generators and
defining relations) of $\fpo(2n)$ (problem discussed
in sec.~1.7)
as of a {\it Lie} algebra, see \cite{LP}; for superizations and
open problems, see \cite{GLP}.

3) For $q$-quantization of the finite dimensional Poisson
Lie
superalgebras, see \cite{LSa}; one can also try to derive the
$q$-quantum version of defining relations of the Poisson Lie
superalgebras given in \cite{GLP}.

\section{Deformations
of  the Buttin superalgebra and its
subalgebras}

For preliminaries and definitions, see Appendix: background.

\ssec{2.1.  The main deformation} (After \cite{ALSh}.)  As is
clear from the definition of the Buttin bracket, see (\ref{eq76}),
there is a regrading (namely, $\fb (n; n)$, given by $\deg\xi_i=0,
\deg q_i=1$ for all $i$) under which $\fb(n)$, initially of depth
2, takes the form $\fg=\mathop{\oplus}\limits_{i\geq -1}\fg_i$
with $\fg_0\simeq\fvect(0|n)$ and $\fg_{-1}\cong \Pi(\Cee[\xi])$.
Now, let us replace the $\fvect(0|n)$-module $\fg_{-1}$ of
functions (with inverted parity) with the module of
$\lambda$-densities, i.e., set $\fg_{-1}\cong
\Pi(\Vol(0|n)^\lambda)$, where the action of $\fg_0=\fvect(0|n)$
is given for any $D\in\fg_{0}$ and $f\in \Cee[\xi]$ by the
formulas
\begin{equation}
\label{eq6}
L_D(fvol_\xi^\lambda)=\left (D(f)+(-1)^{p(D)p(f)}\lambda f\Div
D\right )\cdot vol_\xi^\lambda\; \text{
and
$p(vol_\xi^\lambda)=\od$}.
\end{equation}
Define $\fb_{\lambda}(n; n)$, a deform of $\fb(n; n)$, as the {\it Cartan
prolong}
\begin{equation}
\label{eq7}
\fb_{\lambda}(n; n):=(\fg_{-1},
\fg_0)_*=(\Pi(\Vol(0|n)^\lambda),
\fvect(0|n))_*.
\end{equation}
These $\fb_{\lambda}(n; n)$ for all $\lambda$'s constitute the
{\it main deformation}.  (Though main, this deformation is not the
quantization of the Buttin bracket, cf.  sec.~2.2.)

The deform $\fb_{\lambda}(n)$ of $\fb(n)$ is a regrading of
$\fb_{\lambda}(n; n)$ described as follows. Let $\lambda
=\frac{2a}{n(a-b)}$; set
\begin{equation}
\label{eq8}
\fb_{a, b}(n)
=\{M_f\in \fm (n)\mid a\;
\Div M_f=(-1)^{p(f)}2(a-bn)\pderf{f}{\tau}\}.
\end{equation}
Taking into account the explicit form (\ref{eq74}) of the
divergence of $M_{f}$ we see that
\begin{equation}
\label{eq9}
\renewcommand{\arraystretch}{1.4}
\begin{array}{l}
    \fb_{a, b}(n) =\{M_f\in \fm (n)\mid (bn-aE)\pderf{f}{\tau} =a\Delta
f\}=\\
\{D\in\fvect(n|n+1) \mid L_{D}(\vvol_{q,
\xi,
\tau}^a\alpha_{0}^{a-bn})=0\}.
\end{array}
\end{equation}
It is subject to a direct check that $\fb_{a, b}(n)$ is another
notation for $\fb_\lambda(n)$, where $\lambda =\frac{2a}{n(a-b)}$.
This
shows that $\lambda$ actually runs over the projective line
$\Cee P^1$, not $\Cee$.

Observe the following isomorphisms:
\begin{equation}
\label{eq10}
\fb_{nb, b}(n)\cong \fsm(n);\quad \fb_{a, -a}(2; 2)\cong
\fb_{1/2}(2;
2)\cong\fh(2|2), \text{ and }\fb_{-a, -b}(n)\cong \fb_{a, b}(n).
\end{equation}
Moreover, $\fb_\lambda(2; 2)\simeq\fh_\lambda(2|2)$, where
$\fh_\lambda(2|2)$, the deform of $\fh(2|2)$, is described in
sec.~3.2.

The Lie superalgebra $\fb(n)=\fb_{0}(n)$ is not simple: it has an
$\eps$-dimensional, i.e., $(0|1)$-dimensional, center.  At $\lambda=1$
and $\infty$
the Lie superalgebra $\fb_{\lambda}(n)$ is not simple
either: it has a simple ideal of codimension $\eps^n$ and $\eps^{n+1}$,
respectively, cf. \cite{LS1}.  The corresponding exact
sequences are
\begin{equation}
\label{eq11}
\renewcommand{\arraystretch}{1.4}
\begin{array}{c}
0\tto \Cee \cdot M_{1} \tto \fb(n)\tto
\fle(n)\tto 0,\\
0\tto \fb_{1} \degree(n)\tto \fb_{1}(n)\tto \Cee\cdot
M_{\xi_1\dots\xi_n}
\tto 0,\\
0\tto \fb_{\infty} \degree(n)\tto \fb_{\infty}(n)\tto \Cee\cdot M_{\tau\xi_1\dots\xi_n}
\tto 0.\\
\end{array}
\end{equation}
Clearly, at the exceptional
values of $\lambda$, i.e., 0, 1, and
$\infty$, the deformations of $\fb_{\lambda}(n)$ should be
investigated extra carefully.  As we will see immediately, it pays: in
each of
exceptional points we find extra deformations.

The Lie superalgebras $\fb_\lambda(n)$ are simple for $n>1$ and
$\lambda\neq 0, 1, \infty$.  It is also clear that
the
$\fb_{\lambda}(n)$ are non-isomorphic for distinct $\lambda$'s for
$n>2$.

\ssec{Grozman's twist of the Schouten bracket} 1) The {\it
Schouten bracket} was originally defined on the superspace of
multivector fields on a manifold, i.e., on the superspace of
sections of the exterior algebra (over the algebra $\cF$ of
functions) of the tangent bundle,
$\Gamma(\Lambda^{\bcdot}(T(M)))\cong\Lambda^{\bcdot}_\cF(Vect(M))$.
The explicit formula of the Schouten bracket (in which the hatted
slot should be ignored, as usual) is
\begin{equation}
\label{eq14}
\renewcommand{\arraystretch}{1.4}
\begin{array}{c}
[X_1\wedge\dots \wedge\dots \wedge X_k, Y_1\wedge\dots
\wedge Y_l]=\\
\sum_{i, j}(-1)^{i+j}[X_i, Y_j]\wedge X_1\wedge\dots\wedge
\hat
X_i\wedge \dots\wedge X_k\wedge Y_1\wedge\dots\wedge
\hat Y_j\wedge
\dots \wedge Y_l.
\end{array}
\end{equation}
With the help of Sign Rule we easily superize formula (\ref{eq14})
for the case when $M$ is replaced with a supermanifold $\cM$.  The
relation of the superversion of (\ref{eq14}) thus obtained with
(\ref{eq66}) is as follows. Let $x$ and $\xi$ be the even and odd
coordinates on $\cM$.  Setting $\theta_i=\Pi(\pder{x_{i}})=\check
x_{i}$, $q_j=\Pi(\pder {\xi_{j}})=\check \xi_{j}$ we get an
identification of the Schouten bracket of multivector fields on
$\cM$ with the Buttin bracket of functions on the supermanifold
$\check\cM$ whose coordinates are $x, \xi$ and $\check x$, $\check
\xi$; any transformation of $x, \xi$ induces that of the checked
coordinates.

2) In \cite{G}, Grozman classified all bilinear invariant differential
operators acting in the spaces of sections of tensor fields on
any
manifold.  In this remarkable paper, he also introduced a
one-parameter deformation of the Schouten bracket related with the one
we call ``main deformation''. Namely, he introduced
the
operator
\begin{equation}
\label{eq15}
\renewcommand{\arraystretch}{1.4}
\begin{array}{l}
    X\vvol^\mu, Y\vvol^\nu \mapsto \left ((\nu-1)(\mu +
    \nu -1) \Div X \cdot Y + \right .\\
(-1)^{p(X)}(\mu-1)(\mu
+ \nu -1)X \Div Y -\\
\left. (\mu- 1)(\nu -1)
\Div(XY)\right )\vvol^{\mu+ \nu},\end{array}
\end{equation}
where the {\it divergence of a polyvector field} is best described
in local coordinates $(x, \check x)$ on the supermanifold $\check
M$ associated with any supermanifold $M$, see formula
(\ref{eq76}).

Grozman's Lie superalgebra on twisted polyvector fields on $M$
given by formula (\ref{eq18}) can be realized as a subalgebra of
the Lie superalgebra of divergence-free polyvector fields
$\fs\fb(n+1)$ of on $M\times \Ree_{+}$, or the Lie subsuperalgebra
of functions (with respect to the Buttin bracket) on the
associated supermanifold with checked coordinates.  The exact
formula:
\begin{equation}
\label{eq16} X\vvol^{\lambda}\longmapsto
t^{-\lambda}X+\frac{1}{\lambda-1}t^{-\lambda+1} \pder{t}\wedge
\Div (X);
\end{equation}
in terms of sec.~A.4 the right hand side of (\ref{eq16}) is
$t^{-\lambda}f(x, \xi)+ \frac{1}{\lambda-1}t^{-\lambda+1} \check
{t}\Delta (f)$.

\ssec{The case $n=2$} Let $\bar n$ denote the grading
$$
\text{$\deg q_{i}=0$, $\deg \xi_{i}=0$ for $i=1, \dots , n$.}
$$
Then we have an analog of representation (\ref{eq7}):
\begin{equation}
\label{eqb} \fg_i=\left(\Pi(\Lambda
^{i+1}(\fvect(n|0))\right)\otimes \Vol^{-i\lambda}\;\text{ for
$i=-1, 0, \dots , n-1$.}
\end{equation}
Since
\begin{equation}
\label{eqvolbar} \Lambda^n(\fvect(n|0))\simeq\Vol^{-1}(n|0),
\end{equation}
we see that if $\lambda=-\frac12$, then
$$
\fg_{-1}\simeq\fg_{1}\simeq\Vol^{-1/2}.
$$
Generally,
$$
\fb_{\lambda}(2; \bar 2)\simeq \fb_{-1-\lambda}(2; \bar 2)\text{
or, which is the same, $\fh_{\lambda}(2|2; \bar 2)\simeq
\fh_{-1-\lambda}(2|2; \bar 2)$}.
$$
In particular, we have an additional outer automorphism $T_{\pm}:
\fg_{-1}\longleftrightarrow \fg_1$ of $\fg=\fb_{-1/2}(2; \bar 2)$.

Now recall (see Background) that the natural symmetric paring
$$
(f\sqrt{\vvol}, g\sqrt{\vvol})=\int fg\vvol
$$
(well defined on functions with compact support and formally
extended to formal semi-densities) becomes {\bf skew} symmetric on
the purely odd space, and hence determines a central extension.
This is the well-known extension that determines the Poisson
superalgebra; this cocycle is of degree $-2$.

Now observe that (compare with (\ref{eqvolbar}))
\begin{equation}
\label{eqvvol} \Lambda^n(\fvect(0|n))\simeq\Vol(0|n)
\end{equation}
which leads to the isomorphism
$$
\fb_{\lambda}(2; 2)\simeq \fb_{1-\lambda}(2; 2), \text{ or, which
is the same, $\fh_{\lambda}(2|2)\simeq \fh_{1-\lambda}(2|2)$}.
$$
In particular, (this is the trifle Kochetkov, and all of us,
missed):
$$
\fb_{1/2}(2; 2)\simeq \fb_{-3/2}(2; 2).
$$

\ssec{2.2.  Quantizations: retelling \cite{Ko1}, \cite{Ko2}, \cite{L3}
and more} The deformation $\fb_{\lambda}(n)$ of $\fb(n)$ that
connects
$\fb(n)$ with $\fsm(n)$ will be referred to as the {\it main} one.
The other deformations, called {\it singular} ones, are no less
interesting.  Of particular interest are the
ones corresponding to
$\lambda=0$ and (for $n=2$) to $\lambda=\frac12$ and
$\lambda=-\frac32$: they are {\it quantizations}.

For $\fg=\fb_{\lambda}(n)$, set $H=H^2(\fg; \fg)$.
(Recall (see
\cite{F}) that the superspace $H$ is usually identified with the
tangent space to the singular supervariety of parameters of
deformations of $\fg$ at the point
corresponding to $\fg$.)

\begin{Theorem} {\em 1)} $\dim ~H=(1|0)$ for
$\fg=\fb_{\lambda}(n)$ unless $\lambda=0$, $-1$, $1$, $\infty$ for
$n>2$.  For $n=2$, in addition to the above $\dim
~H\neq (1|0)$ at
$\lambda=\frac12$.

{\em 2)} At exceptional values of $\lambda$ listed in $1)$ we have

$\dim ~H=(2|0)$ at $\lambda=\pm 1$ and $n$ odd, or $\lambda=\infty$
and $n$
even, or $\lambda=\frac12$ (or $\lambda=-\frac32$) and $n=2$.

$\dim ~H=(1|1)$ at $\lambda=0$, or $\lambda=\infty$ and $n$ odd, or
$\lambda=\pm 1$ and $n$ even.

\noindent The
corresponding cocycles $C$ are given by the following nonzero
values in terms of the generating functions $f$ and $g$, where
$d_{\od}(f)$ is the degree of $f$ with respect to odd
indeterminates
only (here $k=(k_{1}, \dots , k_{n})$; we set $q^k=q_{1}^{k_{1}}\dots
q_{n}^{k_{n}}$ and $|k|=\sum k_{i}$):
\begin{equation}
\label{eq17}
\begin{tabular}{|c|c|c|}

\hline
$\fb_{\lambda}(n)$&$p(C)$&$C(f, g)$\cr
\hline
$\fb_{0}(n)$&odd&$(-1)^{p(f)}(d_{\od}(f)-1)(d_{\od}(g)-1)fg$\cr
\hline
$\fb_{-1}(n)$&$n+1\pmod 2$&$f=q^k, \,g=q^l \mapsto
(4-|k|-|l|)q^{k+l}\xi_{1}\dots
\xi_{n}+$\cr
&&$\tau \Delta(q^{k+l}\xi_{1}\dots \xi_{n})$\cr
\hline
$\fb_{1}(n)$&$n+1\pmod 2$&$f=\xi_{1}\dots \xi_{n}, \,g\mapsto
\begin{cases}(d_{\od}(g)-1)g&\text{if $g\neq af$,
$a\in\Cee$}\\
 2(n-1)f &\text{if $g=f$ and $n$ is even}\end{cases}$
 \cr
\hline
$\fb_{\infty}(n)$&$n\pmod 2$&$f=\tau\xi_{1}\dots \xi_{n}, \,g\mapsto
\begin{cases}(d_{\od}(g)-1)g&\text{if
$g\neq af$, $a\in\Cee$}\\
 2f &\text{if $g=f$ and $n$ is odd}\end{cases}$
 \cr
\hline
\end{tabular}
\end{equation}
On $\fb_{\frac12}(2)\simeq\fh(2|2; 1)$ (the latter being a
regrading of $\fh(2|2)$, see sec.~A.7) the cocycle is the one
induced on
 $\fh(2|2)=\fpo(2|2)/\text{center}$ by the usual quantization of
 $\fpo(2|2)$: we first quantize $\fpo$ and then take
the quotient
 modulo center (generated by constants).

{\em 3)} The space $H$ is diagonalizable with respect to the Cartan
subalgebra of $\fder ~\fg$; the cocycle $M$ corresponding
to the main
deformation is one of the eigenvectors.  Let $C$ be another
eigenvector in $H$, it determines a singular deformation.  The only
cocycles $kM+lC$ that can be extended to
a global deformation are
those for $kl=0$,i.e., either $M$ or $C$.

All the singular deformations of the bracket $\{\cdot, \cdot\}_{old}$
in $\fb_{\lambda}(n)$ (except the one for
$\lambda=\frac12$ and $n=2$)
have a very simple form even for the even $\hbar$:
\begin{equation}
\label{eq18}
\{f, g\}_{\hbar}^{sing}=\{f, g\}_{old}+\hbar\cdot C(f, g)\text{ for any } f,
g\in
\fb_{\lambda}(n).
\end{equation}
\end{Theorem}

\begin{Remark} C.~Roger observed that the singular deformation
(quantization) of $\fb_{0}(n)=\fb(n)$ is, up to sign, the wedge product
of two 1-cocycles, the
derivations $f\mapsto (d_{\od}(f)-1)f$.  He
also advises to note that the cocycle on
$\fb_{\frac12}(2)\simeq\fh(2|2; 1)$ induced by the quantization of
$\fpo(2|2)$ is a
straightforward superization of the well-known Vey's cocycle \cite{GS}.
\end{Remark}

Since the elements of $\fb_{\lambda}(n)$ are encoded by functions (for
us: polynomials) in $\tau$, $q$ and
$\xi$ subject to one relation with
an odd left hand side in which $\tau$ enters, it seems plausible that
the bracket in $\fb_{\lambda}(n)$ can be, at least for generic values
of
parameter $\lambda$, expressed solely in terms of $q$ and $\xi$.
Indeed, here is the explicit formula (in which $\{f, g\}_{B.B.}$ is the
usual
antibracket):
\begin{equation}
\label{eq19}
\renewcommand{\arraystretch}{1.4}
\begin{array}{l}
    \{f_1, f_2\}_{\lambda}^{main}=\{f_1, f_2\}_{B.B.}+\lambda(c_{\lambda}(f_1,
f_2)f_1\Delta f_2 + (-1)^{p(f_1)}c_{\lambda}(f_2,
f_1)(\Delta
f_1)f_2),
\end{array}
\end{equation}
where
\begin{equation}
\label{eq20}
\displaystyle c_{\lambda}(f_1, f_2)=\frac{\deg
f_1-2}{2+\lambda(\deg f_2 -n)}
\end{equation}
and $\deg$ is computed with respect
to the standard grading $\deg
q_{i}=\deg \xi_{i}=1$.

\section{The main deformation of $\fh(2|2)$}

Comparison of the non-positive terms of the $\Zee$-gradings shows
that $\fb_{\lambda}(2; 2)\cong
\fh_{\lambda}(2|2)$.

In Eq.  (\ref{eq9}) we have interpreted $\fb_{\lambda}(n)$ as
preserving a complicated tensor $\vvol_{q, \xi,
\tau}^a\alpha_{0}^{a-bn}$.

\ssbegin{3.1}{Theorem}
Set
$\hbar(\lambda)=\frac{2\lambda-1}{\lambda}$.  Then
$D\in\fvect(2|2)$ belongs to $\fb_{\lambda}(2; 2)$ if and only if
\begin{equation}
\label{eq21}
D=D_{f}=H_{f}+\hbar(\lambda)W_{f},\text{ where
$W_{f}=\left (\int_{0}^p
\pder{\eta}\pderf{f}{\xi}dp\right )\partial_{p}+
(-1)^{p(f)}\pderf{f}{\xi}\partial_{\eta}$}
\end{equation}
for some $f\in \Cee[p, q, \xi, \eta]$. Then, for $f, g\in \Cee[p,
q, \xi, \eta]$, we
have
\begin{equation}
\label{eq22}
[D_{f}, D_{g}]=D_{\{f, g\}_{P.B.}}+\hbar(\lambda)D_{c(f, g)},
\end{equation}
where
\begin{equation}
\label{eq23}
\renewcommand{\arraystretch}{1.4}
\begin{array}{l}
    c(f,
g)=-\displaystyle\pderf{f}{p}\int_{0}^p\pder{\eta}\pderf{g}{\xi}
dp+ \pderf{g}{p}\int_{0}^p\pder{\eta}\pderf{f}{\xi}
dp+\\
\displaystyle\pder{\eta}\big(\int_{(0, q)}^{(p, q)}\left(
(-1)^{p(f)}\pderf{f}{p}\pderf{g}{\xi}-\pderf{f}{\xi}\pderf{g}{p}\right)dp+\\
\displaystyle\int_{(0, 0)}^{(0, q)}\left(
(-1)^{p(f)}\pderf{f}{q}\pderf{g}{\xi}-
\pderf{f}{\xi}\pderf{g}{q}\right)|_{p=0}dq\big)+\\
\displaystyle\xi\pder{\xi}\left(
(-1)^{p(f)}\pderf{f}{\xi}\pderf{g}{\eta}+
\pderf{f}{\eta}\pderf{g}{\xi}\right)|_{p=0, q=0}.
\end{array}
\end{equation}
\end{Theorem}

\noindent Observe that the formula
\begin{equation}
\label{eq24}
[H_f, H_g]_{new}=H_{\{f, g\}_{P.B.}}+\hbar
(\lambda)\cdot H_{c(f, g)}
\end{equation}
determines a deformation of $\fh(2|2)$ (which is the main deformation
of $\fb_{1/2}(2)$) but (and this agrees with \cite{BT1}) the formula
\begin{equation}
\label{eq25}
\{f,
g\}_{new}=\{f, g\}_{P.B.}+\hbar (\lambda)\cdot c(f, g)
\end{equation}
does not determine a deformation of $\fpo(2|2)$ because (\ref{eq25}) does
not satisfy the Jacobi identity.

\ssec{3.2.
Deformations of $\fg=\fb_{1/2}(n; n)$}
Clearly, $\fg_{-1}$ is isomorphic to $\Pi(\sqrt{Vol})$. Therefore
there is an embedding
\begin{equation}
\label{eq26}
\fb_{1/2}(n;
n)\subset\begin{cases}\fh(2^{n-1}|2^{n-1})&\text{ for
$n$
even}\\
\fle(2^{n-1})&\text{ for $n$ odd.}\end{cases}
\end{equation}
It is tempting to determine quantizations of $\fg$ in addition to
those considered by Kochetkov, as the composition of embedding
(\ref{eq29})  and the subsequent quantization.

For $n=2$, when (\ref{eq29}) is not just an embedding but an
isomorphism, this certainly works and we get the following extra
quantization of the antibracket described in Theorem 1.2: we first
deform  the antibracket to the point $\lambda =\frac12$ along the
main deformation, and then quantize it as the quotient of the
Poisson superalgebra. This scheme fails to give new algebras for
$n=2k>2$:

\begin{Theorem} For $n=2k>2$, the image of $\fb_{1/2}(n; n)$
under
embedding (\ref{eq26}) is rigid under the quantization of the ambient.
\end{Theorem}

Proof: direct verification.

\ssec{3.3.  General algebras are rigid} The rigidity of contact
and pericontact series was earlier established by painstaking
calculations due to Shmelev \cite{Sm} for the series $\fk$ and
Kochetkov \cite{Ko2} for the series $\fm$.  It is, however, an
example of the general statements on cohomology of coinduced
modules (\cite{F}) and immediately follows from the later
computations of cohomologies of $\fgl(m|n)$, $\fosp(m|2n)$, and
$\fpe(n)$, see \cite{FL} and \cite{F}, and the following
observation: as modules over themselves, the algebras
$\fg=\fvect$, $\fk$ and $\fm$ are expressed as modules of
generalized tensor fields (see sec.~A.6) as follows:
\begin{equation}
\label{eq27}
\fvect(m|n)=T(\id_{\fgl(m|n)}); \quad
\fk(2m+1|n)=T(\Cee[-2]_{\fg_{0}});\quad \fm(n)=\Pi(T(\Cee[-2]_{\fg_{0}})),
\end{equation}
where $\Cee[k]$ is
the representation of $\fg_{0}$ (in the
standard grading of $\fg$) trivial on the simple part and such
that the center $z$ of $\fg_{0}$ acts as multiplication by
$k\in\Cee$, where
the central element $z$ is selected to act on
$\fg_{i}$ as multiplication by $i\in\Zee$.

Thus, the adjoint modules are coinduced, and therefore we have:

\begin{Theorem} $H^2(\fg;
\fg)\simeq H^2(\fg_{0}; \id_{\fg_{0}})=0$
for $\fg=\fvect(m|n)$, and $H^2(\fg; \fg)\simeq H^2(\fg_{0};
\Cee[-2]_{\fg_{0}})=0$ for $\fk(2m+1|n)$ and
$\fm(n)$.
\end{Theorem}

\section{Representations of $\fhei(2n|m)$ and $\fab(n)$}

We begin with observation that only for the standard realization
(see sec.~A.7) the relation between the
commutation/anticommutation relations (represented by the elements
of negative degree from the Poisson or Buttin Lie superalgebra)
and the Poisson or Buttin Lie superalgebra itself is the same as
for Lie algebra $\fpo(2n)$.  (To see the difference most
graphically, consider the finite dimensional case, say,
$\fpo(0|2n)$ with the grading $\deg\xi_{i}=0$, $\deg\eta_{i}=1$
for all $i$.)

\ssbegin{4.1}{Lemma} {\em (\cite{Sg1}) 1)} Let $V$ be a vector
superspace, and $P_V(C)=(V\otimes C)_\ev$ be the set of its
$C$-points.  Then $V\simeq W$ if
and only if $P_V(C)\simeq P_W(C)$ for
all supercommutative superalgebras $C$.

{\em 2)} Let $\fg$ and $\fh$ be Lie superalgebras.  Then
$\fg\simeq\fh$ if and only if
$P_\fg(C)\simeq P_\fh(C)$ as Lie
algebras for all supercommutative superalgebras $C$.

{\em 3)} Let $V$ and $W$ be two modules over $\fg$.  The modules are
isomorphic if and only if
$P_V(C)\simeq P_W(C)$ as modules over
$P_\fg(C)$ for all supercommutative superalgebras $C$.

{\em iv)} It suffices to verify the above conditions for $C=\Lambda(N)$
with $N$ ``sufficiently
large''.
\end{Lemma}

\begin{Remark} One should not replace $N$ with $\infty$, as Berezin
did: though we only have to verify one condition instead of infinitely
many ones, we
acquire infinite topological difficulties, cf.
\cite{D2}. \end{Remark}

\ssec{4.2.  Irreducible representations of $\fhei(2n|m)$ and its
analogs} The following statement and its analog,
heading 1) of
Theorem 4.3, are particular case of a result of Sergeev \cite{Sg2}
(that corrects ``Theorem'' 7 of \cite{K}):

\begin{Theorem} Let us represent the superspace of
$\fhei(2n|m)$ as $W\oplus \Cee z$, where $W$ is endowed with the
form $B$, see $\ref{eq80}$ and $\ref{eq82}$, and represent $W$ as
$V\oplus V^*$ if $m=2k$ or $W=V\oplus V^*\oplus U$ if $m=2k+1$,
where $V$ and $V^*$ are isotropic with respect to the form $B$ and
each of dimension $n|k$.  Then over $\Cee$, the only irreducible
representations of $\fhei(2n|m)$ are isomorphic to the following
Fock superspaces: $\cF_{\hbar}\simeq\Cee[V]$ if $m=2k$ and this is
a $G$-type representation; or $\cF_{\hbar}\simeq\Cee[V\oplus U]$
if $m=2k+1$ and this is a Q-type representation, i.e.,
$\fhei(2n|m)$ maps into $\fq(V\oplus U)$.  On $\cF_{\hbar}$, the
center $z$ of $\fhei(2n|m)$ acts as a scalar multiplication by
$\hbar$.
\end{Theorem}

\ssec{4.3.  Irreducible representations of $\fab(n)$} Recall, see
sec.~ A.5.2, that $\fab(n)=W\oplus \Cee z$.  Let $W$ be spanned by
the even elements $q_1$, \dots , $q_n$ and odd elements
$\theta_1$, \dots , $\theta_n$.

\begin{Theorem} {\em 1)} Over any commutative algebra with the zero
odd part, $\fab(n)$ has only two irreducible representations:
the
$1|0$-dimensional trivial module ${\bf 1}$ and $\Pi({\bf 1})$.

{\em 2)} Let $C$ be a supercommutative superalgebra with
$C_\od\neq 0$ and $\xi\in C_\od$.  There are $n+1$
distinct
irreducible $\fab(n; C)$-modules $\cF_i$, $0\leq i\leq n$,
corresponding to odd parameters describing the tangent space to
the trivial representation ${\bf 1}$. Namely, for a nonzero
vector
$v$, set $zv=\xi v$ and
\begin{equation}
\label{eq28}
q_1v=\dots =q_iv=\theta_{i+1}v=\dots =\theta_{n}v=0 \text{ for $i=0, 1,
\dots, n$}
\end{equation}
and
define
\begin{equation}
\label{eq29}
\renewcommand{\arraystretch}{1.4}
\begin{array}{l}
    \cF_i=\ind_{\Span (q_1, \dots , q_i, \theta_{i+1}, \dots , \theta_{n};
z)}^{\fab(n)}(Cv)\simeq \\
C[\lambda_1, \dots , \lambda_i; x_{i+1}, \dots
,
x_{n}].\end{array}
\end{equation}
The explicit realization of the operators is:
\begin{equation}
\label{eq30}
\renewcommand{\arraystretch}{1.4}
\begin{array}{l}
    q_1\mapsto\xi\pder{\lambda_1}, \dots ,
q_i\mapsto\xi\pder{\lambda_i};
q_{i+1}\mapsto x_{i+1},
\dots , q_{n}\mapsto x_{n}\\
\theta_{1}\mapsto  \lambda_1, \dots ,\theta_{i}\mapsto  \lambda_i;
\theta_{i+1}\mapsto -\xi\pder{x_{i+1}}, \dots ,
\theta_{n}\mapsto
-\xi\pder{x_{n}}.
\end{array}
\end{equation}
\end{Theorem}

Thus, {\it as superspaces},
\begin{equation}
\label{eq31}
\renewcommand{\arraystretch}{1.4}
\begin{array}{l}
    \fpo(2n|2m)\simeq\Cee[q,
\pder{q}, \xi,
\pder{\xi}]\simeq U(\fhei(2n|2m))/(z-\hbar);\\
\Pi(\fb(n)\otimes C[\xi])\simeq
\left(U(\fab(n))\otimes C[\xi]\right)/(z-\xi),
\end{array}
\end{equation}
where for the
antibracket we have to consider everything over $C$ to
account for the odd parameters.

\section{Appendix: Background}

\ssec{A.1.  Linear algebra in superspaces.  Generalities} A
{\it
superspace} is a $\Zee /2$-graded space; for any superspace
$V=V_{\ev}\oplus V_{\od }$, denote by $\Pi (V)$ another copy of
the same superspace: with the shifted parity,
i.e.,
$(\Pi(V))_{\bar i}= V_{\bar i+\od }$.  The {\it superdimension} of
$V$ is $\dim V=p+q\eps $, where $\eps ^2=1$ and $p=\dim V_{\ev}$,
$q=\dim V_{\od }$. Usually, $\dim V$ is expressed
as a pair $p|q$;
in this notation the useful formula $\dim V\otimes W=\dim V\cdot
\dim W$ looks mysterious whereas with $\eps$ this is clear.

A superspace structure in $V$ induces
the superspace structure in
the space $\End (V)$.  A {\it basis of a superspace} is always a
basis consisting of {\it homogeneous} vectors.  Let $p_i$ denote
the parity of $i$th
basis vector, then the matrix unit $E_{ij}$ is
supposed to be of parity $p_i+p_j$ and the bracket of
supermatrices is defined via Sign Rule:

{\it if something of parity $p$ moves
past something of parity $q$ the
sign $(-1)^{pq}$ accrues; the formulas defined on homogeneous elements
are extended to arbitrary ones via linearity}.

More examples of application
of Sign Rule: setting $[X,
Y]=XY-(-1)^{p(X)p(Y)}YX$ we get the notion of the supercommutator and
the ensuing notions of the supercommutative superalgebra and the Lie
superalgebra
(that in addition to superskew-commutativity satisfies
the super Jacobi identity, i.e., the Jacobi identity amended with the
Sign Rule).  The {\it superderivation} of a superalgebra
$A$ is a
linear map $D: A\tto A$ that satisfies the super Leibniz rule
\begin{equation}
\label{eq32}
D(ab)=D(a)b+(-1)^{p(D)p(a)}aD(b).
\end{equation}
In particular, let $A=\Cee[x]$ be the free supercommutative
polynomial
superalgebra in $x=(x_{1}, \dots , x_{n})$, where the superstructure
is determined by the parities of the indeterminates: $p(x_{i})=p_{i}$.
Partial derivatives are defined (with
the help of super Leibniz Rule)
by the formulas $\pderf{x_{i}}{x_{j}}=\delta_{i,j}$.  Clearly, the
collection $\fder A$ of all superderivations of $A$ is a Lie
superalgebra whose
elements are of the form $ \sum
f_i(x)\pder{x_{i}}$.

We consider the exterior differential as a superderivation of the
superalgebra of exterior forms, so $dx$ is even for any odd
$x$ and we
can consider not only polynomials in $dx$.  Smooth or analytic
functions in $dx$ are called {\it pseudodifferential forms} on the
supermanifold with coordinates $x$, see
\cite{BL}.  We needed them
to interpret $\fh_{\lambda}(2|2)$.

\ssec{A.1.1.  General linear superalgebras: two types} The {\it
general linear} Lie superalgebra of all
supermatrices of given
format $\Par$ (an ordered collection of parities of basis vectors,
or just a superdimension) is denoted by $\fgl(\Par)$.  Any matrix
from $\fgl(\Par)$ can be expressed
as the sum of its even and odd
parts; in the standard (simplest) format this is the following
block expression:
\begin{equation}
\label{eq33}
\begin{pmatrix}A&B\\
C&D\end{pmatrix}=\begin{pmatrix}A&0\\
0&D\end{pmatrix}+\begin{pmatrix}0&B\\ C&0\end{pmatrix},\quad
p\left(\begin{pmatrix}A&0\\
0&D\end{pmatrix}\right)=\ev, \; p\left(\begin{pmatrix}0&B\\
C&0\end{pmatrix}\right)=\od.
\end{equation}
The
{\it supertrace} is the map $\fgl (\Par)\tto \Cee$,
$(A_{ij})\mapsto \sum (-1)^{p_{i}}A_{ii}$.  Since $\str [x, y]=0$, the
subsuperspace of supertraceless matrices constitutes the
{\it special
linear} Lie subsuperalgebra $\fsl(\Par)$.

Another super versions of $\fgl(n)$ is called the {\it queer} Lie
superalgebra and is defined as the Lie superalgebra that
preserves the
complex structure given by an {\it odd} operator $J$, i.e., is the
centralizer $C(J)$ of $J$:
\begin{equation}
\label{eq34}
\fq(n)=C(J)=\{X\in\fgl(n|n)\mid [X, J]=0 \}, \text{ where
}
J^2=-\id.
\end{equation}
It is clear (over $\Cee$) that by a change of basis we can reduce $J$
to the form $J_{2n}=\begin{pmatrix}0&1_n\\ -1&0\end{pmatrix}$.  In the
standard format we
have
\begin{equation}
\label{eq35}
\fq(n)=\left \{\begin{pmatrix}A&B\\ B&A\end{pmatrix}\right\}.
\end{equation}
On $\fq(n)$, the {\it queertrace} is defined: $\qtr:
\begin{pmatrix}A&B\\
B&A\end{pmatrix}\mapsto \tr B$.  Denote by $\fsq(n)$
the Lie
superalgebra of {\it queertraceless} matrices.

Observe that the identity representations of $\fq$ and $\fsq$ in
$V$, though irreducible in super setting, are not irreducible in
the non-graded sense: take homogeneous (with respect to parity)
and linearly independent vectors $v_1$, \dots , $v_n$ from $V$;
then $\Span (v_1+J(v_1), \dots , v_n+J(v_n))$ is an invariant
subspace of $V$ which is not a subsuperspace. On such
inhomogeneous irreducible representations, see \cite{Lan}.

A representation is {\it irreducible} \index{representation of Lie
superalgebra irreducible} \index{$G$-type
irreducible representation
of Lie superalgebra} of {\it general type} or just of {\it $G$-type}
if there is no nontrivial invariant subspace.  An irreducible
representation is called {\it
irreducible of $Q$-type} \index{$Q$-type
irreducible representation of Lie superalgebra} ($Q$ is after the
general queer Lie superalgebra); if it has no invariant
sub{\it
super}space but {\it has} a nontrivial invariant subspace.

\ssec{A.1.2.  Lie superalgebras that preserve bilinear forms: two
types} Given a linear map $F$ of superspaces, there exists
a
corresponding dual map $F^*$ between the dual superspaces; if $A$ is
the supermatrix corresponding to $F$ in a basis of format $\Par$, then
the {\it supertransposed} matrix $A^{st}$
corresponds to $F^*$:
\begin{equation}
\label{eq36}
(A^{st})_{ij}=(-1)^{(p_{i}+p_{j})(p_{i}+p(A))}A_{ji}.
\end{equation}

The supermatrices $X\in\fgl(\Par)$ such that
\begin{equation}
\label{eq37}
X^{st}B+(-1)^{p(X)p(B)}BX=0\quad \text{for an
homogeneous matrix
$B\in\fgl(\Par)$}
\end{equation}
constitute the Lie superalgebra $\faut (B)$ that preserves the
bilinear form on $V$ with matrix $B$.

Recall that the {\it supersymmetry} of
the homogeneous form $\omega $
means that its matrix $B$ satisfies the condition $B^{u}=B$, where
\begin{equation}
\label{eq38}
B^{u}=
\begin{pmatrix}
R^{t} & (-1)^{p(B)}T^{t} \\ (-1)^{p(B)}S^{t}
&
-U^{t}\end{pmatrix}\text{
for the matrix $B=\begin{pmatrix} R& S \\ T & U\end{pmatrix}$. }
\end{equation}
Similarly, {\it skew-supersymmetry} of $B$ means that $B^{u}=-B$.
Thus, we see that the {\it
upsetting} of bilinear forms $u:\Bil
(V, W)\tto\Bil(W, V)$, which, for the {\it spaces} $V=W$, is
expressed on matrices in terms of the transposition, becomes a new
operation on
supermatrices.

The most popular canonical forms of the nondegenerate supersymmetric
form are the ones whose supermatrices in the standard format are the
following canonical ones,
$B_{ev}$ or $B'_{ev}$:
\begin{equation}
\label{eq39}
B_{ev}(m|2n)= \begin{pmatrix}
1_m&0\\
0&J_{2n}
\end{pmatrix},\quad \text{where
$J_{2n}=\begin{pmatrix}0&1_n\\-1_n&0\end{pmatrix}$},
\end{equation}
or
\begin{equation}
\label{eq40}
B'_{ev}(m|2n)=
\begin{pmatrix}
\antidiag (1, \dots , 1)&0\\
0&J_{2n}
\end{pmatrix}.
\end{equation}
The usual notation for $\faut (B_{ev}(m|2n))$ is $\fosp(m|2n)$ or,
more precisely, $\fosp^{sy}(m|2n)$.  Observe
that the passage from $V$
to $\Pi (V)$ sends the supersymmetric forms to superskew-symmetric
forms, preserved by the \lq\lq symplectico-orthogonal" Lie
superalgebra, $\fsp'\fo
(2n|m)$ or, better say, $\fosp^{sk}(m|2n)$,
which is isomorphic to $\fosp^{sy}(m|2n)$ but has a different matrix
realization.  We never use notation $\fsp\fo (2n|m)$ in order
to
prevent confusion with the special Poisson superalgebra.

In the standard format the matrix realizations of these algebras
are:
\begin{equation}
\label{eq41}
\begin{matrix}
\fosp (m|2n)=\left\{\left
(\begin{matrix} E&Y&X^t\\
X&A&B\\
-Y^t&C&-A^t\end{matrix} \right)\right\};\quad \fosp^{sk}(m|2n)=
\left\{\left(\begin{matrix} A&B&X\\
C&-A^t&Y^t\\
Y&-X^t&E\end{matrix} \right)\right\},
\\
\text{where}\;
\left(\begin{matrix} A&B\\
C&-A^t\end{matrix} \right)\in \fsp(2n),\quad E\in\fo(m)\;
\text{and}\;  {}^t \; \text{is the usual transposition}.\end{matrix}
\end{equation}

A
non-degenerate supersymmetric odd bilinear form $B_{odd}(n|n)$
can be reduced to a canonical form whose matrix in the standard
format is $J_{2n}$.  A canonical form of the superskew
odd
non-degenerate form in the standard format is
$\Pi_{2n}=\begin{pmatrix} 0&1_n\\1_n&0\end{pmatrix}$.  Observe
that we did not make a mistake here: with a minus for the
symmetric form
and with a plus for the skew form!

The usual notation for $\faut (B_{odd}(\Par))$ is $\fpe(\Par)$.
The passage from $V$ to $\Pi (V)$ establishes an isomorphism
$\fpe^{sy}(\Par)\cong\fpe^{sk}(\Par)$.  This Lie superalgebra is
called, as A.~Weil suggested, {\it periplectic}\footnote{An ``odd
[analog of] symplectic'' (the orthogonal group preserves lines, as
its name reflects (${\stackrel{,}{o}\rho\theta\acute o\varsigma=}$
straight, direct (the opposite of crooked is
${\eps\stackrel{,}{\upsilon}\theta\acute \upsilon\varsigma}$)),
symplectic (${\sigma\upsilon\mu\pi\lambda}\acute
{\eps\kappa\eps\iota\upsilon})=$ intertwine or interweave, and
${\pi\eps\rho\iota\sigma\sigma\acute o\varsigma}$ means odd, as
opposed to even.}.  The matrix realizations in the standard format
of these superalgebras are:
\begin{equation}
\label{eq42}
\begin{matrix}
\fpe ^{sy}\ (n)=\left\{\begin{pmatrix} A&B\\
C&-A^t\end{pmatrix}, \; \text{where}\; B=-B^t,\;
C=C^t\right\};\\
\fpe^{sk}(n)=\left\{\begin{pmatrix}A&B\\ C&-A^t\end{pmatrix},
\;
\text{where}\; B=B^t,\;  C=-C^t\right\}.
\end{matrix}
\end{equation}
We note that although $\fosp^{sy} (m|2n)\simeq\fosp ^{sk} (2n|m)$,
as well as $\fpe ^{sy} (n)\simeq\fpe ^{sk} (n)$, the difference
between these isomorphic Lie superalgebras is sometimes crucial,
see \cite{LS1} and Remark~A.5.2.

The {\it special periplectic} superalgebra is
$\fspe(n)=\{X\in\fpe(n)\mid
\str X=0\}$.  Of particular interest will be also
$\fspe(n)_{a, b}=\fspe(n)\supplus\Cee(az+bd)$, where $z=1_{2n}$,
$d=\diag(1_{n}, -1_{n})$.  Indeed, it is the linear part of
$\fb_{a,
b}(n)$.

\ssec{A.2.  Vectorial Lie superalgebras.  The standard realization}
The elements of the Lie algebra $\cL=\fder\; \Cee [[u]]$ are
considered as vector fields.
The Lie algebra $\cL$ has only one
maximal subalgebra $\cL_0$ of finite codimension (consisting of the
fields that vanish at the origin).  The subalgebra $\cL_0$ determines
a
filtration of $\cL$: set
\begin{equation}
\label{eq43}
\cL_{-1}=\cL\; \text{ and }\; \cL_i =\{D\in \cL_{i-1}\mid  [D,
\cL]\subset\cL_{i-1}\}\; \text{for }i\geq 1.
\end{equation}
The associated graded Lie algebra
$L=\mathop{\oplus}\limits_{i\geq
-1}L_i$, where $L_i=\cL_{i}/\cL_{i+1}$, consists of the vector fields
with {\it polynomial} coefficients.

\ssec{A.2.1. Superization}  For
$\cL=\fder\, \Cee [u, \xi]$ suppose
$\cL_0\subset\cL$ is a maximal subalgebra of finite codimension and
containing no ideals of $\cL$.  Let $\cL_{-1}$ be a minimal subspace
of $\cL$
containing $\cL_0$, different from $\cL_0$ and
$\cL_0$-invariant.  A {\it Weisfeiler filtration} of $\cL$ is
determined  by setting for $i\geq 1$:
\begin{equation}
\label{eq44}
\cL_{-i-1}=[\cL_{-1},
\cL_{-i}]+\cL_{-i}\; \text{ and }\; \cL_i
=\{D\in \cL_{i-1}\mid  [D, \cL_{-1}]\subset\cL_{i-1}\}.
\end{equation}
Since the codimension of $\cL_0$ is finite, the filtration takes the
form
\begin{equation}
\label{eq45}
\cL=\cL_{-d}\supset\dots\supset\cL_{0}\supset\dots
\end{equation}
for some {\it depth} $d$.  Considering the subspaces (\ref{eq43}) as the
basis of a topology, we can complete the
graded or filtered Lie
superalgebras $L$ or $\cL$; the elements of the completion are the
vector fields with formal power series as coefficients.  Although the
structure of the
graded algebras is easier to describe, in
applications the completed Lie superalgebras are usually needed.

Unlike Lie algebras, simple vectorial {\it super}algebras possess
{\it several} non-isomorphic maximal subalgebras of finite
codimension, see sec.~A.7.

1) {\it General algebras}.  Let $x=(u_1, \dots , u_n, \theta_1, \dots
, \theta_m)$, where the $u_i$ are even
indeterminates and the
$\theta_j$ are odd ones.  Set $\fvect (n|m)=\fder\; \Cee[x]$; it is
called {\it the general vectorial Lie superalgebra}.  \index{$\fvect$
general vectorial
Lie superalgebra}\index{ Lie superalgebra general
vectorial}

On vectorial superalgebras, there are two types of trace. The
divergences (depending on a fixed volume element) belong
to one of
them, various linear functionals that vanish on the brackets
(traces) belong to the other type. Accordingly, the {\it special}
({\it divergence free}) subalgebra of a
vectorial algebra $\fg$ is
denoted by $\fs\fg$, e.g., $\fvect(n|m)$ and $\fsvect(n|m)$, and
the traceless subalgebra of $\fg$ is denoted $\fg'$.

2) {\it Special algebras}.  The {\it
divergence}\index{divergence} of
the field $D=\sum\limits_if_i\pder{u_{i}} + \sum\limits_j
g_j\pder{\theta_{j}}$ is the function (in our case: a polynomial, or a
series)
\begin{equation}
\label{eq46}
\Div
D=\sum\limits_i\pderf{f_{i}}{u_{i}}+
\sum\limits_j (-1)^{p(g_{j})}
\pderf{g_{i}}{\theta_{j}}.
\end{equation}

$\bullet$ The Lie superalgebra $\fsvect (n|m)=\{D \in \fvect
(n|m)\mid
\Div D=0\}$ is called the {\it special} or {\it divergence-free
vectorial superalgebra}.  \index{$\fsvect$ general vectorial Lie
superalgebra}\index{ Lie superalgebra special
vectorial}\index{ Lie
superalgebra divergence-free}

It is clear that it is also possible to describe $\fsvect(n|m)$ as
$\{ D\in \fvect (n|m)\mid  L_D\vvol _x=0\}$, where $\vvol_x$ is
the volume form with constant coefficients in coordinates $x$ (see
sec.~A.6) and $L_D$ the Lie derivative with respect to $D$.

$\bullet$ The Lie superalgebra
$\fsvect_{\lambda}(0|m)=\{D \in \fvect
(0|m)\mid  \Div (1+\lambda\theta_1\cdot \dots \cdot \theta_m)D=0\}$, where
$p(\lambda)\equiv m\pmod 2$, --- the deform of $\fsvect(0|m)$ --- is
called the {\it
deformed special} or {\it divergence-free vectorial
superalgebra}.  Clearly, $\fsvect_{\lambda}(0|m)\cong
\fsvect_{\mu}(0|m)$ for $\lambda\mu\neq 0$.  So we briefly denote
these deforms
by $\widetilde{\fsvect}(0|m)$.

Observe that, for $m$ odd, the parameter of deformation $\lambda$
is odd.

3) {\it The algebras that preserve Pfaff equations and
differential
2-forms}.  Having denoted $u=(t, p_1, \dots , p_n, q_1, \dots , q_n)$
set
\begin{equation}
\label{eq47}
\tilde \alpha_1 = dt +\sum\limits_{1\leq i\leq n}(p_idq_i - q_idp_i)\
+ \sum\limits_{1\leq j\leq
m}\theta_jd\theta_j\quad\text{and}\quad
\omega_0=d\alpha_1\ .
\end{equation}
$\bullet$ The form $\tilde \alpha_1$ is called {\it contact}, the form $\tilde
\omega_0$ is called {\it symplectic}.\index{form
differential
contact}\index{form differential symplectic} Sometimes it is more
convenient to redenote the $\theta$'s and
set
\begin{equation}
\label{eq48}
\xi_j=\frac{1}{\sqrt{2}}(\theta_{j}-i\theta_{r+j});\quad \eta_j=\frac{1}{
\sqrt{2}}(\theta_{j}+i\theta_{r+j})\; \text{ for}\; j\leq r= [m/2]\; (\text{here}\;
i^2=-1),
\quad
\theta =\theta_{2r+1}
\end{equation}
and in place of $\tilde \omega_0$
or $\tilde \alpha_1$ take $\alpha_1$
and $\omega_0=d\alpha_1$, respectively, where
\begin{equation}
\label{eq49}
\alpha_1=dt+\sum\limits_{1\leq i\leq n}(p_idq_i-q_idp_i)+
\sum\limits_{1\leq j\leq
r}(\xi_jd\eta_j+\eta_jd\xi_j)
\left\{\begin{matrix}&
\text{ if }\ m=2r\\
+\theta d\theta&\text{ if }\ m=2r+1.\end{matrix}\right.
\end{equation}

The Lie superalgebra that preserves the {\it Pfaff
equation}
\index{Pfaff equation} $\alpha_1(X)=0$ for $x\in \fvect(2n+1|m)$,
i.e., the superalgebra
\begin{equation}
\label{eq50}
\fk (2n+1|m)=\{ D\in \fvect (2n+1|m)\mid L_D\alpha_1=f_D\alpha_1\text{
for some
}f_D\in \Cee [t, p, q, \theta]\},
\end{equation}
is called the {\it contact superalgebra}.\index{$\fk$ contact
superalgebra} \index{Lie superalgebra contact} The Lie superalgebra
\begin{equation}
\label{eq51}
\fpo (2n|m)=\{
D\in \fk (2n+1|m)\mid  L_D\alpha_1=0\}
\end{equation}
is called the {\it Poisson} superalgebra.\index{$\fpo$ Poisson
superalgebra} (A geometric interpretation of the Poisson superalgebra:
it
is the Lie superalgebra that preserves the connection with form
$\alpha$ in the line bundle over a symplectic supermanifold with the
symplectic form $d\alpha$.)

$\bullet$
Similarly, set $u=q=(q_1, \dots , q_n)$,
let $\theta=(\xi_1, \dots , \xi_n; \tau)$ be odd. Set
\begin{equation}
\label{eq52}
\alpha_0=d\tau+\sum\limits_i(\xi_idq_i+q_id\xi_i), \qquad\qquad
\omega_1=d\alpha_0
\end{equation}
and
call these forms the {\it pericontact} and {\it periplectic},
respectively.\index{form differential pericontact}\index{form
differential periplectic} Observe that this pericontact
form is
even.

The Lie superalgebra that preserves the Pfaff equation
$\alpha_0(X)=0$ for $x\in \fvect(n|n+1)$, i.e., the superalgebra
\begin{equation}
\label{eq53}
\fm (n)=\{ D\in \fvect (n|n+1)\mid
L_D\alpha_0=f_D\cdot \alpha_0\text{ for
some }\; f_D\in \Cee [q, \xi, \tau]\}
\end{equation}
is called the {\it pericontact superalgebra}.\index{$\fm$
pericontact superalgebra}

The Lie superalgebra
\index{$\fb$ Buttin superalgebra}
\begin{equation}
\label{eq54}
\fb (n)=\{ D\in \fm (n)\mid  L_D\alpha_0=0\}
\end{equation}
is referred to as the {\it Buttin} superalgebra.  (A geometric
interpretation of the Buttin
superalgebra: it is the Lie superalgebra
that preserves the connection with form $\alpha_1$ in the line bundle
of rank $\varepsilon$ over a periplectic supermanifold, i.e.,
a
supermanifold with the periplectic form $d\alpha_0$.)

The Lie superalgebras
\begin{equation}
\label{eq55}
\fsm (n)=\{ D\in \fm (n)\mid  \Div\ D=0\}\ , \; \fs\fb (n)=\{ D\in
\fb (n)\mid \Div\ D=0\}
\end{equation}
are called the
{\it divergence-free} (or {\it special}) {\it
pericontact} and {\it special Buttin} superalgebras, respectively.

\begin{rem*}{Remark} A relation with finite dimensional geometry
is as
follows.  Clearly, $\ker \alpha_1= \ker \tilde\alpha_1$.  The
restriction of $\tilde \omega_0$ to $\ker \alpha_1$ is the
orthosymplectic form $B_{ev}(m|2n)$; the restriction
of $\omega_0$ to
$\ker \tilde \alpha_1$ is $B'_{ev}(m|2n)$.  Similarly, the restriction of
$\omega _1$ to $\ker \alpha_0$ is $B_{odd}(n|n)$.
\end{rem*}

\ssec{A.3. Generating
functions} A laconic way to describe
$\fk$, $\fm$ and their subalgebras is via generating functions.

$\bullet$ Odd form $\alpha_1$.  For $f\in\Cee [t, p, q, \theta]$,
we
set\index{$K_f$ contact vector field} \index{$H_f$ hamiltonian
vector field}:
\begin{equation}
\label{eq56}
K_f=(2-E)(f)\pder{t}-H_f +
\pderf{f}{t} E,
\end{equation}
where $E=\sum\limits_i y_i \pder{y_{i}}$ (here the
$y_{i}$ are all the
coordinates except $t$) is the {\it Euler operator} (which counts the
degree with respect to the $y_{i}$), and $H_f$ is the hamiltonian field
with Hamiltonian
$f$ that preserves $d\tilde \alpha_1$:
\begin{equation}
\label{eq57}
H_f=\sum\limits_{i\leq n}\left(\pderf{f}{p_i}
\pder{q_i}-\pderf{f}{q_i} \pder{p_i}\right)
-(-1)^{p(f)}\left(\sum\limits_{j\leq m}\pderf{
f}{\theta_j}
\pder{\theta_j}\right ) .
\end{equation}

The choice of the form $\alpha_1$ instead of $\tilde\alpha_1$ only
affects the shape of $H_f$ that we give for
$m=2k+1$:
\begin{equation}
\label{eq58}
H_f=\sum\limits_{i\leq n} \left(\pderf{f}{p_i}
\pder{q_i}-\pderf{f}{q_i} \pder{p_i}\right)
-(-1)^{p(f)}\sum\limits_{j\leq k}\left(\pderf{f}{\xi_j}
\pder{\eta_j}+ \pderf{f}{\eta_j}
\pder{\xi_j}+ \pderf{f}{\theta}
\pder{\theta}\right).
\end{equation}

$\bullet$ Even form $\alpha_0$. For $f\in\Cee [q, \xi, \tau]$, we
set:
\begin{equation}
\label{eq59}
M_f=(2-E)(f)\pder{\tau}- Le_f
-(-1)^{p(f)}
\pderf{f}{\tau} E,
\end{equation}
where $E=\sum\limits_iy_i
\pder{y_i}$ (here the $y_i$ are all the coordinates except
$\tau$), and
\begin{equation}
\label{eq60}
Le_f=\sum\limits_{i\leq n}\left(
\pderf{f}{q_i}\
\pder{\xi_i}+(-1)^{p(f)} \pderf{f}{\xi_i}\
\pder{q_i}\right).
\end{equation}
\index{$M_f$ contact vector field} \index{$Le_f$ periplectic vector
field}

Since
\begin{equation}
\label{eq61}
\renewcommand{\arraystretch}{1.4}
\begin{array}{l}
L_{K_f}(\alpha_1)=2 \pderf{f}{t}\alpha_1=K_1(f)\alpha_1, \\
L_{M_f}(\alpha_0)=-(-1)^{p(f)}2
\pderf{
f}{\tau}\alpha_0=-(-1)^{p(f)}M_1(f)\alpha_0,
\end{array}
\end{equation}
it follows that $K_f\in \fk (2n+1|m)$ and $M_f\in \fm (n)$. Observe that
\begin{equation}
\label{eq62}
p(Le_f)=p(M_f)=p(f)+\od.
\end{equation}

$\bullet$ To the (super)commutators $[K_f,
K_g]$ or $[M_f, M_g]$ there
correspond {\it contact brackets}\index{Poisson bracket}\index{contact
bracket} of the generating functions:
\begin{equation}
\label{eq63}
[K_f, K_g]=K_{\{f, \;
g\}_{k.b.}};\quad\quad [M_f, M_g]=M_{\{f, \;
g\}_{m.b.}}
\end{equation}
The explicit formulas for the contact brackets are as follows.  Let us
first define the brackets on functions that do not depend on
$t$
(resp.  $\tau$).

The {\it Poisson bracket} $\{\cdot , \cdot\}_{P.b.}$ (in the realization with the form
$\omega_0$) is given by the formula
\begin{equation}
\label{eq64}
\{f, g\}_{P.b.}=\sum\limits_{i\leq n}\
\bigg(\pderf{f}{p_i}\
\pderf{g}{q_i}-\ \pderf{f}{q_i}\
\pderf{g}{p_i}\bigg)-(-1)^{p(f)}\sum\limits_{j\leq m}\
\pderf{f}{\theta_j}\ \pderf{g}{\theta_j}\text{ for }f, g\in \Cee [p,
q,
\theta]
\end{equation}
and in the realization with the form
$\omega_0$ for $m=2k+1$ it is given by the formula
\begin{equation}
\label{eq65}
\renewcommand{\arraystretch}{1.4}
\begin{array}{l}
    \{f,
g\}_{P.b.}=\sum\limits_{i\leq n}\ \bigg(\pderf{f}{p_i}\
\pderf{g}{q_i}-\ \pderf{f}{q_i}\
\pderf{g}{p_i}\bigg)-\\
(-1)^{p(f)}\bigg[\sum\limits_{j\leq m}( \pderf{f}{\xi_j}\
\pderf{
g}{\eta_j}+\pderf{f}{\eta_j}\ \pderf{ g}{\xi_j})+\pderf{f}{\theta}\
\pderf{ g}{\theta}\bigg]\text{ for }f, g\in \Cee [p, q, \xi, \eta, \theta].
\end{array}
\end{equation}

The {\it Buttin bracket} $\{\cdot , \cdot\}_{B.b.}$ \index{Buttin
bracket $=$ Schouten bracket} is given by the formula
\begin{equation}
\label{eq66}
\{ f, g\}_{B.b.}=\sum\limits_{i\leq n}\
\bigg(\pderf{f}{q_i}\
\pderf{g}{\xi_i}+(-1)^{p(f)}\ \pderf{f}{\xi_i}\
\pderf{g}{q_i}\bigg)\text{ for }f, g\in \Cee [q, \xi].
\end{equation}

In terms of the Poisson and Buttin brackets, respectively, the
contact
brackets are
\begin{equation}
\label{eq67}
\{ f, g\}_{k.b.}=(2-E) (f)\pderf{g}{t}-\pderf{f}
{t}(2-E) (g)-\{ f, g\}_{P.b.}
\end{equation}
and
\begin{equation}
\label{eq68}
\{ f, g\}_{m.b.}=(2-E)
(f)\pderf{g}{\tau}+(-1)^{p(f)}
\pderf{f}{\tau}(2-E) (g)-\{ f, g\}_{B.b.}.
\end{equation}

The Lie superalgebras of {\it Hamiltonian fields}\index{Hamiltonian
vector fields} (or {\it Hamiltonian
superalgebra}) and its
special subalgebra (defined only if $n=0$) are
\begin{equation}
\label{eq69}
\fh (2n|m)=\{ D\in \fvect (2n|m)\mid \ L_D\omega_0=0\}\; \text{ and} \;
\fsh (m)=\{H_f\in \fh (0|m)\mid  \int
f\vvol_{\theta}=0\}.
\end{equation}
The ``odd'' analogs of the Lie superalgebra of hamiltonian fields
are the Lie superalgebra of vector fields $\Le_{f}$ introduced in
\cite{L1} and its special subalgebra:
\begin{equation}
\label{eq70}
\fle
(n)=\{ D\in \fvect (n|n)\mid  L_D\omega_1=0\} \; \text{ and} \;
\fsle (n)=\{ D\in \fle (n)\mid  \Div D=0\}.
\end{equation}

It is not difficult to prove the following isomorphisms (as
superspaces):
\begin{equation}
\label{eq71}
\renewcommand{\arraystretch}{1.4}
\begin{array}{rclrcl}
\fk (2n+1|m)&\cong&\Span(K_f\mid  f\in \Cee[t, p, q, \xi]);&\fle
(n)&\cong&\Span(Le_f\mid  f\in \Cee [q, \xi]);\\
\fm
(n)&\cong&\Span(M_f\mid  f\in \Cee [\tau, q, \xi]);&
\fh (2n|m)&\cong&\Span(H_f\mid  f\in
\Cee [p, q, \xi]).
\end{array}
\end{equation}

\ssec{A.4.  Divergence-free subalgebras} Since
\begin{equation}
\label{eq72}
\Div
K_f =(2n+2-m)K_1(f),
\end{equation}
it follows that the divergence-free subalgebra of the contact Lie
superalgebra either coincides with it (for $m=2n+2$) or is
isomorphic to the Poisson
superalgebra.  For the pericontact
series, the situation is more interesting: the divergence free
subalgebra is simple.

Since
\begin{equation}
\label{eq73}
\Div M_f =(-1)^{p(f)}2\left ((1-E)\pderf{f}{\tau} -
\sum\limits_{i\leq
n}\frac{\partial^2 f}{\partial q_i \partial\xi_i}\right ),
\end{equation}
it follows that
\begin{equation}
\label{eq74}
\fsm (n) = \Span\left (M_f \in \fm (n)\mid
(1-E)\pderf{f}{\tau}
=\sum\limits_{i\leq n}\frac{\partial^2 f}{\partial q_i
\partial\xi_i}\right ).
\end{equation}

In particular,
\begin{equation}
\label{eq75}
\Div Le_f = (-1)^{p(f)}2\sum\limits_{i\leq n}\frac{\partial^2 f}{\partial
q_i
\partial\xi_i}.
\end{equation}
The odd analog of the Laplacian, namely, the operator
\begin{equation}
\label{eq76}
\Delta=\sum\limits_{i\leq n}\frac{\partial^2 }{\partial
q_i
\partial\xi_i}
\end{equation}
on a periplectic supermanifold appeared in physics under the name of
{\it BRST operator}, cf.  \cite{GPS}.  The vector fields
from $\fsle (n)$ are generated by {\it harmonic}
functions, i.e., such
that $\Delta(f)=0$.

\ssec{A.5.  The Cartan prolongs} To define $\fb_\lambda(n)$, one
of our main characters, and several related algebras we need the
notion of the
Cartan prolong.  So let us recall the definition and
generalize it somewhat.  Let $\fg$ be a Lie algebra, $V$ a
$\fg$-module, $S^i$ the operator of the $i$th symmetric power.
Set
$\fg_{-1} = V$, $\fg_0 = \fg$ and, for $i > 0$, define the $i$th
{\it Cartan prolong} (the result of Cartan's {\it prolongation})
of the pair $(\fg_{-1}, \fg_0)$
as
\begin{equation}
\label{eq77}
\renewcommand{\arraystretch}{1.4}
\begin{array}{l}
\fg_i = \{X\in \Hom(\fg_{-1}, \fg_{i-1})\mid  X(v_{0})(v_{1}, ..., v_{i})
=\\
X(v_{1})(v_{0}, ..., v_{i})\; \text{ for any }\; v_{0}, v_{1},
...,
v_{i}\in \fg_{-1}\}
= (S^i(\fg_{-1}^*)\otimes \fg_0)\cap (S^{i+1}(\fg_{-1}^*)\otimes
\fg_{-1}).
\end{array}
\end{equation}
(Here we consider $\fg_0$ as a subspace in
$\fg_{-1}^{*}\otimes
\fg_{-1}$, so the intersection is well-defined.)

The {\it Cartan prolong} of the pair $(V, \fg)$ is $(\fg_{-1},
\fg_{0})_* = \mathop{\oplus}\limits_{i\geq -1} \fg_i$.

Suppose
that the $\fg_0$-module $\fg_{-1}$ is {\it faithful}.  Then,
clearly,
\begin{equation}
\label{eq78}
\renewcommand{\arraystretch}{1.4}
\begin{array}{l}
(\fg_{-1}, \fg_{0})_*\subset \fvect (n) = \fder~
\Cee[x_1,
... , x_n],\; \text{ where }\; n = \dim~ \fg_{-1}\; \text{ and }\\
\fg_i = \{D\in \fvect(n)\mid  \deg D=i, [D, X]\in\fg_{i-1}\text{ for any }
X\in\fg_{-1}\}.
\end{array}
\end{equation}
It can
be easily verified that the Lie algebra structure on
$\fvect (n)$ induces same on $(\fg_{-1}, \fg_{0})_*$.

Of the four simple vectorial Lie algebras, three are Cartan
prolongs:
$\fvect(n)=(\id, \fgl(n))_*$, $\fsvect(n)=(\id, \fsl(n))_*$ and
$\fh(2n)=(\id, \fsp(n))_*$. The fourth one --- $\fk(2n+1)$ --- is the
result of a trifle more general construction
described as follows.

\ssec{A.5.1. A generalization of the Cartan prolong} Let
$\fg_-=\mathop{\oplus}\limits_{-d\leq i\leq -1}\fg_i$ be a
nilpotent $\Zee$-graded Lie algebra and
$\fg_0\subset \fder_0\fg$
a Lie subalgebra of the $\Zee$-grading-preserving derivations.
For $i >0$, define the $i$-th prolong of the pair $(\fg_{-},
\fg_0)$ as
\begin{equation}
\label{eq79}
\fg_i =
((S^{\bcdot}(\fg_-^*)\otimes \fg_0)\cap
(S^{\bcdot}(\fg_-^*)\otimes \fg_-))_i,
\end{equation}
where the subscript $i$ in the right hand side singles out the
component of degree $i$.

Define $(\fg_-,
\fg_0)_*$, to be
$\mathop{\oplus}\limits_{i\geq -d}
\fg_i$; then, as is easy to verify, $(\fg_-, \fg_0)_*$ is a Lie algebra.

What is the Lie algebra of contact vector fields in these
terms?
Denote by $\fhei(2n)$ the Heisenberg Lie algebra: its space is
$W\oplus {\Cee}\cdot z$, where $W$ is a $2n$-dimensional space
endowed with a non-degenerate skew-symmetric
bilinear form $B$ and
the bracket in $\fhei(2n)$ is given by the following relations:
\begin{equation}
\label{eq80}
\text{$z$ is even and lies in the center and $[v, w]=B(v, w)\cdot z$ for any $v,
w\in
W$.}
\end{equation}
Clearly, $ \fk(2n+1)\cong (\fhei(2n), \fc\fsp(2n))_*$.

\ssec{A.5.2.  Lie superalgebras of vector fields as Cartan
prolongs} The superization of the constructions from sec.~ A.5 are
straightforward: via Sign Rule.  We thus obtain:
\begin{equation}
\label{eq81}
\renewcommand{\arraystretch}{1.4}
\begin{array}{l}
\fvect(m|n)=(\id, \fgl(m|n))_*; \; \fsvect(m|n)=(\id,
\fsl(m|n))_*; \\
    \fh(2m|n)=(\id, \fosp^{sk}(m|2n))_*; \\
\fle(n)=(\id, \fpe^{sk}(n))_*; \;
\fs\fle(n)=(\id, \fspe^{sk}(n))_*.
\end{array}
\end{equation}
{\it Remark}. Observe that the Cartan
prolongs $(\id, \fosp^{sy} (m|2n))_*$
and $(\id, \fpe ^{sy} (n))_*$ are finite dimensional.

The generalization of Cartan prolongations described in sec.~
A.5.1 has, after superization, {\bf two} analogs associated with
the contact series $\fk$ and $\fm$, respectively.

$\bullet$ Let (\ref{eq80}) define the bracket in the Lie
superalgebra $\fhei(2n|m)$ or $\fhei(W)$ on the direct sum of a
$(2n, m)$-dimensional superspace $W$ endowed with a non-degenerate
skew-symmetric bilinear form $B$ and the $(1, 0)$-dimensional
space spanned by $z$.

Clearly, we have $\fk(2n+1|m)=(\fhei(2n|m), \fc\fosp^{sk}(m|2n))_*$.
More generally, given $\fhei(2n|m)$ and a subalgebra $\fg$ of
$\fc\fosp^{sk}(m|2n)$, we call $(\fhei(2n|m), \fg)_*$
the {\it
$k$-prolong} of $(W, \fg)$, where $W$ is the identity
$\fosp^{sk}(m|2n)$-module.

$\bullet$ The ``odd'' analog of $\fk$ is associated with the
following ``odd'' analog of
$\fhei(2n|m)$.  Denote by $\fab(n)$ or
$\fab(W)$ the {\it antibracket} Lie superalgebra: its space is
$W\oplus \Cee\cdot z$, where $W$ is an $n|n$-dimensional
superspace endowed
with a non-degenerate skew-symmetric odd
bilinear form $B$; the bracket in $\fab(n)$ is given by the
following relations:
\begin{equation}
\label{eq82}
\text{$z$ is odd and lies in
the center; $[v, w]=B(v,
w)\cdot z$ for $v, w\in W$.}
\end{equation}

Clearly, $\fm(n)=(\fab(n), \fc\fpe^{sk}(n))_*$.  More generally, given
$\fab(n)$ and a subalgebra $\fg$ of $\fc\fpe^{sk}(n)$, we call
$(\fab(n),
\fg)_*$ the {\it $m$-prolong} of $(W, \fg)$, where $W$ is
the identity $\fpe^{sk}(n)$-module.

\ssec{A.6.  The modules of tensor fields} To advance further, we
have to recall the
definition of the modules of tensor fields over
$\fvect(m|n)$ and its subalgebras, see \cite{BL}, \cite{L2}. For
any other $\Zee$-graded vectorial Lie superalgebra, the
construction is
identical.

Let $\fg=\fvect(m|n)$ and $\fg_{\geq}=\mathop{\oplus}\limits_{i\geq
0}\fg_{i}$.  Clearly, $\fvect_0(m|n)\cong \fgl(m|n)$.  Let $V$ be the
$\fgl(m|n)$-module with the
{\it lowest} weight $\lambda=\lwt(V)$.
Make $V$ into a $\fg_{\geq}$-module setting $\fg_{+}\cdot V=0$ for
$\fg_{+}=\mathop{\oplus}\limits_{i> 0}\fg_{i}$.  Let us realize $\fg$
by
vector fields on the $m|n$-dimensional linear supermanifold
$\cC^{m|n}$ with coordinates $x=(u, \xi)$.  The superspace
$T(V)=\Hom_{U(\fg_{\geq})}(U(\fg), V)$ is isomorphic, due to
the
Poincar\'e--Birkhoff--Witt theorem, to ${\Cee}[[x]]\otimes V$.  Its
elements have a natural interpretation as formal {\it tensor fields of
type} $V$.  When $\lambda=(a, \dots ,
a)$ we will simply write $T(\vec
a)$ instead of $T(\lambda)$.  We will usually consider $\fg$-modules
induced from irreducible $\fg_0$-modules.

{\it Examples}: $\fvect(m|n)$ as
$\fvect(m|n)$- and
$\fsvect(m|n)$-modules is $T(\id)$.  Further examples: $T(\vec 0)$ is
the superspace of functions; $\Vol(m|n)=T(1, \dots , 1; -1, \dots ,
-1)$ (the semicolon
separates the first $m$ ``even'' coordinates of
the weight with respect to the matrix units $E_{ii}$ of $\fgl(m|n)$
from the ``odd'' coordinates)
is the superspace of {\it densities} or
{\it volume forms}.  We denote
the generator of $\Vol(m|n)$ corresponding to the ordered set of
coordinates $x$ by $\vvol(x)$.  The space of $\lambda$-densities
is
$\Vol^{\lambda}(m|n)=T(\lambda, \dots , \lambda; -\lambda, \dots ,
-\lambda)$; we denote its generator by $\vvol^{\lambda}(x)$.  In
particular, $\Vol^{\lambda}(m|0)=T(\vec \lambda)$
but
$\Vol^{\lambda}(0|n)=T(\overrightarrow{-\lambda})$.

\footnotesize

{\it Remark}.  To view the volume element as \lq\lq $d^mud^n\xi$" is
totally wrong: the Berezinian (superdeterminant) can
never appear as a factor under
the changes of variables.  One can try to use the usual notations of
differentials provided {\it all} the differentials anticommute.  Then
the linear
transformations that do not intermix the even $u$'s with
the odd $\xi$'s multiply the volume element $\vvol(x)$, viewed as the
fraction $\frac{du_1\cdot ...\cdot du_m}{d\xi_1\cdot
...\cdot
d\xi_n}$, by the Berezinian of the transformation.  But how could we
justify this?  Let $x=(u, \xi)$.  If we consider the usual, exterior,
differential forms, then the
$dx_{i}$'s super anti-commute, hence, the
$d\xi_i$ commute; whereas if we consider the {\it symmetric} product
of the differentials, as in the metrics, then the
$dx_{i}$'s
supercommute, hence, the $du_i$ commute.  However, the $\pder{\xi_i}$
anticommute and, from transformations' point of view,
$\pder{\xi_i}=\frac{1}{d\xi_{i}}$.  The notation, $du_1\cdot
...\cdot
du_m\cdot\pder{\xi_1}\cdot \ldots \cdot \pder{\xi_n}$, suggested by
V.~Ogievetsky, is, nevertheless, still wrong: almost any
transformation $A: (u, \xi)\mapsto (v, \eta)$ sends
$du_1\cdot
...\cdot du_m\cdot\pder{\xi_1}\cdot ...\cdot \pder{\xi_n}$ to the
correct element, $\ber (A)(du^m\cdot\pder{\xi_1}\cdot ...\cdot
\pder{\xi_n})$, plus extra terms.
Indeed, the fraction $du_1\cdot
...\cdot du_m\cdot\pder{\xi_1}\cdot ...\cdot \pder{\xi_n}$ is the
highest weight vector of an {\it indecomposable} $\fgl(m|n)$-module
and $\vvol(x)$ is
the notation of the image of this vector in the
1-dimensional quotient module modulo the invariant submodule that
consists precisely of all the extra terms.

\normalsize

\ssec{A.7.
 Nonstandard realizations}

\footnotesize
\begin{equation}
\label{eq84}
\renewcommand{\arraystretch}{1.3}
\begin{tabular}{|c|c|}
\hline
Lie superalgebra & its $\Zee$-grading \\
\hline $\fvect (n|m; r)$, & $\deg u_i=\deg \xi_j=1$  for any $i,
j$ \hskip 4.5 cm
$(*)$\\
\cline{2-2}
$ 0\leq r\leq m$ &
$\deg \xi_j=0$ for $1\leq j\leq r;$\\
&$\deg u_i=\deg \xi_{r+s}=1$ for any $i, s$ \\
\hline
 & $\deg \tau=2$, $\deg q_i=\deg \xi_i=1$  for any $i$ \hskip 3 cm
$(*)$\\
\cline{2-2}
$\fm(n; r),$& $\deg \tau=\deg q_i=1$, $\deg \xi_i=0$ for any $i$\\
\cline{2-2}
$\; 0\leq r\leq n$& $\deg \tau=\deg q_i=2$, $\deg \xi_i=0$ for $1\leq i\leq r
<n$;\\
$r\neq n-1$& $\deg
u_{r+j}=\deg \xi_{r+j}=1$ for any $j$\\
\hline
\cline{2-2}
$\fk (2n+1|m; r)$, & $\deg t=2$,\\
& $\deg p_i=\deg q_i= \deg \xi_j=\deg \eta_j=\deg \theta_k=1$ for
any $i, j, k$\quad
$(*)$ \\
\cline{2-2}
$0\leq r\leq [\frac{m}{2}]$& $\deg t=\deg \xi_i=2$, $\deg
\eta_{i}=0$ for $1\leq i\leq r\leq [\frac{m}{2}]$; \\
$r\neq k-1$ for $m=2k$ and
$n=0$&$\deg p_i=\deg q_i=
\deg \theta_{j}=1$ for $j\geq 1$ and all $i$\\
\hline
$\fk(1|2m; m)$ & $\deg t =\deg \xi_i=1$, $\deg

\eta_{i}=0$ for $1\leq i\leq m$ \\
\hline
\end{tabular}
\end{equation}
\normalsize

For the reasons why $r$ can not take value $n-1$ for $\fm(n)$ and
$k-1$ for $\fk(1|2k)$, irrelevant in this paper but vital in other
problems, we refer the reader to
\cite {LS1}.

{\bf Comments}: The gradings in the series $\fvect$ induce the
gradings in the series $\fsvect$; the gradings in $\fm$ induce the
gradings in $\fb_{\lambda}$, $\fle$, $\fsle$, $\fb$, $\fs\fb$; the
gradings in $\fk$ induce the gradings in $\fpo$, $\fh$.

In (\ref{eq84}) we consider $\fk (2n+1|m)$ as preserving the Pfaff
equation $\alpha(X) =0$ for $X\in \fvect(2n+1|m)$, where
\begin{equation}
\label{eq83} \alpha =dt+\mathop{\sum}\limits_{i\leq
n}(p_idq_i-q_idp_i)+\mathop{\sum}\limits_{j\leq r}
(\xi_jd\eta_j+\eta_jd\xi_j)+\mathop{\sum}\limits_{k\geq
m-2r}\theta_kd\theta_k.
\end{equation}

The standard realizations correspond to $r=0$, they are marked by
$(*)$.  Observe that the codimension of ${\cal L}_0$ attains its
minimum in the standard realization.

\ssec{Acknowledgements} We are thankful: For financial support to
NFR and RFBR grant 99-01-00245, respectively; to Yu.~Kochetkov for
help; to I.~Batalin, P.~Grozman,
Yu.~Kochetkov, A.~Sergeev and
I.~Tyutin and L.~Vaksman for helpful discussions; to M.~Marinov,
S.~Sternberg and M.~Vasiliev for inspiring questions; to Andrej
Kirillov and Bill
Everett for most careful and friendly editing
that immensely improved the text.

\end{document}